\documentclass[12pt,fleqn]{article}
\usepackage{amsmath}
\usepackage{latexsym}
\usepackage{amsfonts}
\usepackage{epsfig}
\newlength{\dinwidth}
\newlength{\dinmargin}
\newcommand{\resection}[1]{\setcounter{equation}{0}\section{#1}}
\newcommand{\resub}[2]{\subsection{#1}}
\newcommand{\appsectio}[1]{\setcounter{section}{0}
         \addtocounter{section}{1} \setcounter{equation}{0}
                         \section*{#1}}
\newcommand{\appsection}[1]{\addtocounter{section}{1} \setcounter{equation}{0}
                         \section*{#1}}
\renewcommand{\theequation}{\thesection.\arabic{equation}}
\newcommand{\f}[2]{\frac{#1}{#2}}

\def\ra{\rightarrow}
\def\dt{\mbox {\boldmath $\Delta$}}

\def\ep{\varepsilon}

\def\k{{\bf k}}
\def\as{\alpha_s}
\def\q{{\bf q}}
\def\FF{{\cal F}}
\def\LL{{\cal L}}
\def\ss{\sigma}
\def\ssh{\hat{\sigma}}
\def\bks{\!\!\!\!\!\!\!\!\!}
\def\th{\hat{t}}
\def\uh{\hat{u}}
\def\sh{\hat{s}}
\def\o{\omega}
\def\g{\gamma}
\def\G{\Gamma}
\def\op{\o_{\mathbb{P}}} 
\newcommand{\be}{\begin{equation}}
\newcommand{\ee}{\end{equation}}
\newcommand{\bea}{\begin{align}}
\newcommand{\eea}{\end{align}}
\newcommand{\nn}{\nonumber}

\begin{document}

\title{$\k$-Factorization and Small-$x$ Anomalous Dimensions}
\author{G. Camici and M. Ciafaloni\\
{\em Dipartimento di Fisica, Universit\`a di Firenze} \\
{\em and INFN, Sezione di Firenze} \\
{\em Largo E. Fermi, 2 - 50125  Firenze}}
\date{DFF 264/01/97}
\maketitle
\thispagestyle{empty}
\begin{abstract}
We investigate the consistency requirements of the next-to leading BFKL 
equation with the renormalization group, with particular emphasis
on running coupling effects and NL anomalous dimensions.
We show that, despite some model dependence of the bare hard Pomeron,
such consistency holds at leading twist level, provided the effective
variable $\as(t)\log(1/x)$ is not too 
large. We give a unified view of 
resummation formulas for coefficient functions and 
anomalous dimensions in the $Q_0$-scheme and we discuss 
in detail the new one for the 
$q\bar{q}$ contributions to the gluon channel.
\end{abstract}
\begin{center}
\end{center}
\newpage
\setcounter{page}{1}

\resection{Introduction}

The understanding of small-$x$ scaling violations in QCD has become 
increasingly important with the discovery of the steep rise of structure 
functions at HERA\cite{1}. Such rise may be related to the hard 
Pomeron\cite{2} small-$x$ behaviour\cite{3}, but may be also understood
through large scaling violations\cite{4} and may also fit in a fixed order 
perturbative approach\cite{5}.

The above ambiguities are partly unavoidable, because of the lack of 
sufficiently precise \cite{1},
independent measurements of small-$x$ parton densities\cite{6}, but are 
partly due to the lack of complete understanding of scaling 
violations, at theoretical level. In fact, in the small-$x$ region, 
the QCD perturbative series is affected by large logarithms - similarly to
what happens in semihard processes \cite{7} - and becomes dependent on the
effective coupling constant $\alpha_s (t) \log{1/x}$ or $\alpha_s (t)/ 
\omega$, where $t \equiv \log(Q^2/\Lambda^2)$ and $\omega = N-1$ is the
moment index.

Therefore, a hierarchy of perturbative terms is defined in the usual way
\begin{equation}
\gamma (\as, \omega) = \gamma_L \left( \f{\as}{\omega} \right) + \as
\gamma_{NL} \left(\f{\as}{\omega} \right) + NNL-\text{terms}
\end{equation}
for both the singlet anomalous dimension matrix and the coefficient 
functions. Furthermore, since gluons couple to electro-weak 
probes only through
quarks, we actually need leading (L) coefficient functions and up to 
next-to-leading (NL) anomalous dimensions, in order to have a factorization 
scheme independent result for DIS.

This theoretical program has been partly achieved at quark level [8-10],
but is not yet complete at gluon level, despite various efforts
[11-14].
Nevertheless, such completion is needed in order to well understand
scaling violations, and then be able to look at the small-$x$
behaviour of initial data\cite{15} so as 
to find the hard Pomeron rise, if any.

In this paper we wish to present a resummation method which is valid
for all the relevant quantities
up to NL level, and to apply it to coefficients and anomalous dimensions
in the so called $Q_0$-scheme \cite{15}, including the new $q\bar{q}$
contributions to the gluon anomalous dimension, already reported
in a short note \cite{16}. The method is based on the idea of 
$\k$-factorization \cite{8,9}, which is just a way of exploiting the 
consistency of high-energy factorized quantities, which are $\k$-dependent, 
with Renormalization Group (R.G.) factorization, which is not. The 
all-order resummation is obtained by $\k$-integrations of low order 
$\k$-dependent kernels.

Therefore, we shall proceed in two steps. Our first goal is to 
explicitate the NL consistency requirements between
high-energy and collinear factorization with running coupling referring,
for definiteness, to hard processes of DIS type, characterized by 
a probe with a large scale $Q$ and by a target with scale $Q_0 \ll Q$.
This will allow us to obtain well defined resummation formulas in terms of 
kernels of BFKL type \cite{2}, up to NL level.

As a second step, we shall proceed to the study of the NL kernel 
itself, of its eigenvalues, and of the resummed formulas for the $q\bar{q}$
contributions. This study is based on 
the phase-space integration of the squared 
$Q\bar{Q}$ production matrix elements \cite{8}, combined with the virtual 
corrections available in the literature \cite{13,14}.

To start illustrating the first step, 
let us notice that $\k$-factorization is due to the exchange 
of an off-shell gluon of virtuality $t\simeq-\k^2$, which behaves, at high 
energies, as a Regge pole yielding a quasi-constant cross-section.
On the other hand, collinear factorization is due to nearly on-shell quarks
and gluons, providing logarithmic scaling violations.

Therefore, the two kinds of factorization are not related a priori. 
In unbroken gauge theories like QCD, a relationship can be established
because the (Regge) gluon polarization coincides, on shell, 
with one of the collinear polarizations \cite{17}. 
For instance, at one loop level,
Regge behaviour is established, in the gluon channel, by the singular 
behaviour of the DGLAP splitting functions \cite{18}
\begin{equation}
P_{gg}(z)\simeq\f{2C_A}{z},\quad\quad P_{gq}(z)\simeq\f{2C_F}{z},
\end{equation}
or in moment index space $\o=N-1$, by a correspondingly singular anomalous 
dimension matrix \vspace{2 cm}
\begin{equation}
\g_{ab}=\f{\bar{\as}}{\o}
\begin{pmatrix}
0 & 0 \\
\f{C_F}{C_A} & 1
\end{pmatrix}
+{\hat{\g}}_{ab}+O(\o)
\end{equation}
\begin{equation*}
\left(a,b\:=\: q,g,~~~~~~~~~~~~~~~~~~~~~~~~\bar{\as}=\f{\as C_A}{\pi}\right).
\end{equation*}

However, establishing such relationship at next to leading level requires 
the treatment of several subtle points. Firstly,
$\k$-factorization is useful in the gluon channel only, while collinear 
factorization involves both gluons and quark-sea. How to recover from 
the former the two-channel picture of the latter?  

Furthermore, even in the gluon channel, the high energy exchange 
does not yield all collinear singular gluon contributions, some of which 
occur at low energies, as is apparent from Eq. (1.3).
How to relate the two gluon definitions and to get all subleading terms?

Finally, the renormalization group evolution with running coupling has 
been always assumed in order to extract physical predictions from 
$\k$-factorization. Is this behaviour really consistent with 
the BFKL equation and its "hard Pomeron" singularity at NL level?

In Sec. 2 we shall show, on the basis of the work of Fadin, Lipatov
and collaborators [11-14] and of our group [8-10,16], 
that a detailed comparison of high-energy 
and renormalization group is nevertheless possible and provides resummation 
formulas for the NL anomalous dimension matrix (and for the leading 
coefficient functions) in the $Q_0$ factorization scheme.

The basic point is that, as it appears from Refs. [8-16] and Ref. \cite{19}, 
$s$-channel iteration is not yet relevant at NL $\log s$ level, while
$t$-channel iteration is provided by a generalized 
high energy integral equation, in which low energy $\k$-dependent 
kernels occur.
This result, with some modifications, forms the basis for the representation 
of the two-scale problems with anomalous dimensions that we provide here.

Some of the questions asked above do have a tricky answer, however. Firstly, 
even at leading twist level, we find that the renormalization group 
representation with running coupling is valid, for $\o\ll 1$, only in the 
regime of $t=\log\f{Q^2}{\Lambda^2}\gg t_0 > 1$, with
\begin{equation}
b\o t > c
\end{equation}
where the critical constant $c$ is related to the BFKL kernel. This means 
that we have to stay away from a $t$-dependent 
singularity in the $\o$ plane, or in other words, that the 
effective variable $\as(t)\log(1/x)$ should not be too large.

Furthermore, the bare hard Pomeron singularity, which occurs in the 
coefficient function, and ultimately dominates the small-$x$ behaviour, 
turns out to be dependent on the large distance behaviour of the running 
coupling, due to a diffusion process towards low values of $\k^2$. 
Fortunately, this fact does not affects the  R. G. factorization,
valid in the regime (1.4) \cite{20}.

Finally, within the anomalous dimension representation, the two-channel 
picture obtains because of the presence of two distinct 
anomalous dimension eigenvalues $\g_+$, 
$\g_-$, $\g_+$ being the leading one. 
While the NL hierarchy is straightforward for the $\g_+$ contributions, it 
requires careful handling of a whole chain of collinear singular 
kernels for the $\g_-$ contribution, as we shall see.

In other words, comparing the high-energy and R. G. approaches is not 
straightforward at NL level, mostly because they resum the perturbative 
series in different orders.

The contents of the paper are as follows. 
In the next section we set up the NL factorization 
framework with scales $Q$ and $Q_0$ and 
we discuss the structure of its solution in the anomalous dimension regime 
defined before, by singling out running coupling effects, and by discussing
the role of the hard Pomeron singularity and its model dependence \cite{20}.

In the following sections we concentrate on the calculation of the 
anomalous dimensions themselves.
In Sec. 3 we provide the $q\bar{q}$ emission kernel for massive quarks 
and in Sec. 4 we combine it with virtual corrections in the massless limit to
get the (regularized) complete eigenvalue and the ensuing 
resummation formulas.
We discuss our results in Sec. 5 and we provide the details of the 
phase-space integration and of the kernel diagonalization in Appendices A-C.

\newpage
\resection{Next-to-Leading Equation and Anomalous Dimension Regime}
\resub{Gluon density and bare Pomeron with running coupling}

Let us formally define the unintegrated gluon density $\FF_\o^{gA}(\k)$ by 
the high-energy limit of the six-point function in Fig. 1. Such density is a 
function of the moment index $\o=N-1$ and of the momentum transfer $\k^2$ 
of the exchanged gluon, and will eventually be measured by coupling it to 
physical probes, as done later on for the DIS structure functions at scale Q.

We shall assume that the gluon density 
satisfies, at NL level, the BFKL-type equation of Fig. 1, 
whose kernel will be 
written in the form
\begin{align}
K_\o(\k,\k^\prime)&\equiv\f{\bar{\as}(\k^2)}{\o}K(\k,\k^\prime)
=\f{\bar{\as}(\k^2)}{\o}\left(K_0(\k,\k^\prime)+\as K_1(\k,\k^\prime)\right),\\
&~~~~~~~~~~~~~~~~\left(\bar{\as}(\k^2)=
\f{\as(\k^2)C_A}{\pi}=\f{1}{\bar{b}\o\log(\k^2/\Lambda^2)}\equiv
\f{1}{\bar{b}\o t}\right)\nn
\end{align}
where we have factored out the running coupling at the "upper" scale $\k^2$, 
i.e.,
\begin{equation}
\as(\k^2)=\as(\mu^2)-b\as^2(\mu^2)\log\f{\k^2}{\mu^2}+....\quad 
=\left({b}\log\f{\k^2}{\Lambda^2}\right)^{-1}.
\end{equation}

Furthermore $K_0$ represents the leading kernel
\begin{equation}
K_0(\k,\k^\prime)=\left.\f{1}{\q^2}\right|_R=\f{1}{\q^2}-
\delta^2(\q)\int_{\lambda^2}^{\k^2}\f{d^2\q}{\q^2},\quad\quad
~~~~~~~~~~~~~~~(\q=\k-\k^\prime),
\end{equation}
and $K_1$ is the NL one, which can be assumed to be scale invariant and 
contains no high-energy gluon exchange contributions. Its specific form has 
been discussed at length in the literature [11-14,16,19], and its 
$N_f$-dependent part will be explicitated in Sec. 3.

Because of scale invariance, we can define the kernel eigenvalue $\chi(\g)$ 
by the equation
\begin{equation}
\int K(\k,\k^\prime)(\k^\prime)^{\g-1}\f{d^2\k^{\prime 2}}{\pi}=
\chi(\g)(\k^2)^{\g-1}.
\end{equation}
The explicit form of $\chi$ is known at leading level, yielding
\begin{equation}
\chi_0(\g)=2\psi(1)-\psi(\g)-\psi(1-\g),
\end{equation}
and the NL $q\bar{q}$ contribution 
$\chi_1^{q\bar{q}}(\g)$ will be given in Sec. 4, where we also prove the 
factorization of the logarithm in Eq. (2.2).

Let us stress the point that factorizing the 
running coupling at scale $\k^2$ 
is not necessarily an indication of the 
"natural" scale of the problem, which 
rather seems to be $\q^2$, the emitted gluon transverse momentum (see Sec. 3). 
In fact, at NL level, a change of scale in the factorized logarithm of 
Eq. (2.2) can be compensated by a corresponding change of the 
scale-invariant kernel $K_1$. The choice of $\k^2$ allows us to deal  
with a type of equation already studied in 
the literature.

Indeed, it was pointed out long ago \cite{21,22} that by writing the master 
equation
\begin{equation}
\FF_{\o}^{gA}(\k)=k_{\o}^{gA}(\k)+\f{\bar{\as}(\k^2)}{\o}
\int K(\k,\k^\prime)\f{d^2\k^\prime}{\pi}\FF_\o^{gA}(\k^\prime)
\end{equation}
in the $\g$-moments representation
\begin{equation}
\FF_\o^{gA}(\k)=(\k^2)^{-1}\int_{\f{1}{2}-i\infty}^{\f{1}{2}+i\infty}
\f{d\g}{2\pi i}\left(\f{\k^2}{\Lambda^2}\right)^\g f_\o^{gA}(\g),
\end{equation}
we get a differential equation in $\g$, with the formal solution
\begin{equation}
f_\o^{gA}(\g)=\exp\left(-\f{1}{\bar{b}\o}X(\g)\right)\int_\g^\infty
d\g^\prime\left(-\f{d}{d\g^\prime}k_\o^{gA}(\g^\prime)\right)\exp\left[
\f{1}{\bar{b}\o}X(\g^\prime)\right].
\end{equation}
Here we have denoted by $X(\g)$ the primitive of the eigenvalue function, 
i.e.,
\begin{align}
X(\g)=&\int_\f{1}{2}^\g d\g^\prime\chi(\g^\prime),
X_0(\g)=&\int_\f{1}{2}^\g d\g^\prime\chi_0(\g^\prime)=
2\psi(1)\left(\g-\f{1}{2}\right)-\log\f{\G(\g)}{\G(1-\g)}
\end{align}
and we have denoted by $k_{\o}^{gA}(\g)$ the $\g$-moment of the (low-energy)
inhomogeneous term in Eq. (2.6).

Actually, the formal solution (2.8) has to be modified in order to 
deal with the running coupling (2.2) in the small $\k^2$ region around the 
Landau pole.
Cutting off [22-24] or smoothing out such a pole results in modifying the 
large $\g^\prime$ region in the representation (2.8) by adding to the 
inhomogeneous term some $\o$-dependent contribution which 
contains the Pomeron singularity \cite{22}.

In Ref. \cite{20} we have investigated the Green's function of Eq. (2.6) 
in the regime $t\gg t_0\equiv\log(Q_0^2/\Lambda^2)$, where 
$Q_0$ represents 
the scale of the inhomogeneous term and we have shown that it 
factorizes in the form 

\begin{align}
\FF_{\o}^{gA}&=\FF_\o^+(\k)g_\o^A(Q_0)+\text{higher twists},
\quad\quad(t>t_0)\\
\intertext{where}
\FF_\o^+(\k)&=\left(\sqrt{\bar{b}\o}\k^2\right)^{-1}
\int_{\f{1}{2}-i\infty}^{\f{1}{2}+i\infty}
\f{d\g}{2\pi i}\left(\f{\k^2}{\Lambda^2}\right)^\g\exp\left(
-\f{1}{\bar{b}\o}X(\g)\right),\\
\intertext{and}
g_\o^A(\k)&=\f{1}{2}\left(\int_0^{\infty+i\ep}+\int_0^{\infty-i\ep}\right)
d\g^\prime\left[-\f{dk_\o^{gA}(\g^\prime)}{d\g^\prime}\right]
\exp\left(\f{1}{\bar{b}\o}
X(\g^\prime)\right)+\\
&+\text{Pomeron terms.}\nn
\end{align}

We have also found that the position and the structure of the Pomeron 
singularity in Eq. (2.12) is dependent on the smoothing out procedure 
of $\as(t)$, and decouples only in the limit of a hard initial scale 
($t_0\gg 1$), being suppressed, in that case, by (roughly) 
powers of $\Lambda^2/Q_0^2$.

The results above show that, in the limit of $t\gg t_0\gg 1$, we have
exact consistency with the na$\ddot{\imath}$ve perturbative result in Eq. 
(2.8), because we can set $\g\simeq 0$ in the second integral of this 
equation. If instead $t_0$ is of order unity, the structure of the large 
$t$ solution remains factorized, but the coefficient in front is not 
really calculable, being dependent on the low $t$ behaviour of $\as$, i. e. 
on soft hadronic interactions.

In either case the NL BFKL equation remains consistent with the R. G.. 
In fact, in the anomalous dimension regime, defined by
\begin{equation}
\bar{b}\o t>\chi\left(\f{1}{2}\right),
\end{equation}
the large $t$ behaviour of Eq. (2.11) is dominated by the perturbative branch
$\bar{\g}(\as(t))$ of the saddle point equation
\begin{equation}
1=\f{1}{\bar{b}\o t}\chi(\bar{\g}).
\end{equation}

By integrating in $\g$ over the spread 
$|\delta\g|\simeq \sqrt{-\bar{b}\o/\chi^\prime(\bar{\g})}$ 
on the imaginary axis, 
we find the expression 
\begin{equation}
\k^2\FF_\o^{+}(\k)=\f{1}{\sqrt{-\chi^\prime(\bar{\g})}}
\exp{F_\o\left[\bar{\g}(\as(t),t)\right]},
\end{equation}
where the exponent
\begin{equation}
F_\o(\bar{\g},t)=\bar{\g}t-\f{1}{\bar{b}\o}X(\bar{\g})=\int^t\bar{\g}
(\as(\tau))d\tau
\end{equation}
is the one predicted by the R. G..
It is clear that if the limitation (2.14) is not satisfied, the 
representation (2.15) breaks down, due to large $\g$-fluctuations. 

Then, by introducing the integrated gluon density
\begin{equation}
G_\o^{gA}(Q)=\int_0^{Q^2}\f{d^2\k}{\pi}\FF_\o^{gA}(\k)=
\int_{\f{1}{2}-i\infty}^{\f{1}{2}+i\infty}\f{d\g}{2\pi i\g}
e^{\g t}f_\o^A(\g),
\end{equation}
we finally obtain , by Eq. (2.16), the R. G. representation
\begin{equation}
G_\o^{gA}(Q)=H\left(\bar{\g}(\as(t))\right)\exp\left[F_\o\left(\bar{\g}
(\as(t),t)\right)\right]K^A(\o),
\end{equation}
where the l. h. coefficient is given by
\begin{align}
H(\bar{\g})=& \f{1}{\bar{\g}}\f{1}{\sqrt{-\chi^\prime(\bar{\g})}},
\end{align}
while the r. h. coefficient $K^A(\o)$ contains the Pomeron 
singularity and is thus, strictly speaking, not calculable.

Several remarks are in order. First, 
in the case several saddle points are present, a sum over them is understood. 
At NL level, we need at least two of them corresponding to the two eigenvalues
$\g_+$, $\g_-$ (Sec. 2.2).

Secondly, in the limit $Q_0^2\gg\Lambda^2$, the na$\ddot{\imath}$ve 
expression (2.8) applies, and we can set
\begin{equation}
K^A(\o)= \f{\sqrt{\bar{b}\o}}{2\pi}\int_0^\infty d\g^\prime
\left(\f{d}{d\g^\prime}k_\o^A(\g^\prime)\right)\exp\left(\f{1}{\bar{b}\o}
X(\g^\prime)\right),
\end{equation}
so that a perturbative evaluation is possible, as in the following  
examples.

{\em Initial gluon with fixed virtuality.} In this case, we use 
$\k$-factorization at the lower scale also, so that we can fix the gluon 
virtuality in a gauge invariant way, by setting $\k^2=Q_0^2>\Lambda^2$. 
The inhomogeneous term becomes a $\delta$-function in the $\k$-space, 
or 
\begin{equation}
k(\g^\prime)=e^{-\g^\prime t_0}
\end{equation}
in the $\g$-space. For sufficiently large $t_0$ ($\bar{b}\o t_0\gg\chi(1/2)$) 
we can use a saddle point located at $\bar{\g}(\as(t_0))\gg\bar{\g}(\as(t))$ 
at the lower scale also with $\g$-spread 
$|\delta \g|=\sqrt{\f{\bar{b}\o}{-\chi^\prime(\bar{\g}_0)}}$, 
so as to get the approximate 
expression
\begin{equation}
K^g(Q_0)\simeq\f{\bar{b}\o t_0}{\sqrt{-\chi^\prime(\bar{\g}(\as(t_0))}}
\exp\left[-F_\o\left(\bar{\g}(\as(t_o),t_0\right)\right]
\end{equation}
which is $t$-independent as expected.

{\em Initial probe at scale $Q_0$}. This scale can be due, for instance, 
to heavy quark production ($Q_0=M$), or to a virtual electroweak probe. 
In either case one should use $\k$-factorization to define $k^{gA}$ as
an off-shell cross section along the lines of Refs. [8-10].
In all these cases, the $k^{gA}(\g)$ function has, by scale invariance,
a $t_0$-dependence of the type  (2.21) apart from a $\g$-dependent factor, 
computed in some cases \cite{8,17,25}, 
which is analytic in the strip $|Re~\g|<1$. 
The argument about $t$-scale dependence of $K$ goes through as before.

Note, that the perturbative evaluation (2.22) is valid only if the Pomeron 
term in $k^A$ can be neglected. This implies that $\as(t_0)\log(1/x)$ 
should not be too large at the lower scale also. The latter assumption 
is not required in the more general expression (2.18).
\resub{Quark sea and anomalous dimension matrix in $Q_0$-scheme}

It is convenient to introduce the quark-sea contribution 
by the flavour-singlet 
part of the $F_2$ structure function itself (DIS scheme for the quark).
This one can be divided in a direct low-energy 
contribution $G_0$, which contains collinear singularities but not 
$\o=0$ singularities, and in a gluon exchange part (Fig. 2). 
The latter is evaluated by using $\k$-factorization, in terms of the 
unintegrated gluon density defined before, as follows
\begin{equation}
G_\o^{qA}(Q)-G_{0\o}^{qA}(Q)=N_f\int d^2\k\ssh_2(Q,\k,m)\FF^{gA}(\k),
\end{equation}
where $\ssh_2$ is the gauge-invariant off-shell cross-section introduced 
in Ref. \cite{8} and $m$ is the quark mass, that will eventually be set to 
zero.

By introducing the $\g$-representation of Eq. (2.7), the r.h.s. of Eq. (2.23)
takes the form
\begin{equation}
N_f\int\f{d\g}{2\pi i}f_\o^A(\g)e^{\g t}\int\f{d\g_1}{2\pi i}
\left(\f{Q^2}{m^2}\right)^{1-\g-\g_1}H_\o^{ab}(\g,\g_1),
\end{equation}
where we have introduced the double $\g$-moments of $\ssh_2$, that is , 
the abelian contribution to the $q\bar{q}$ $H$-function, already 
computed in Ref. \cite{8} for heavy flavour production. In the $\o=0$ limit,
we have
\begin{align}
H^{ab}(\g_1,\g_2)=&\f{\as}{\pi}\G(1-\g_1-\g_2)\G(\g_1)\G(\g_2)
\left[\f{B(1-\g_1,1-\g_1)B(1-\g_2,1-\g_2)}{4}+\right.\nn\\
&\left.+\f{B(1-\g_1,2-\g_1)B(1-\g_2,2-\g_2)}{(3-2\g_1)(3-2\g_2)}
\left(1+(1-\g_1)(1-\g_2)\right)\right]
\end{align}

The massless quark limit on Eq. (2.24) singles out the pole at $\g_1=1-\g$ 
of the expression (2.25), i.e.
\begin{align}
H_2(\g)=&\underset{\g_1\ra 1-\g}{\lim} H^{ab}(\g_1,\g)(1-\g_1-\g)=\nn\\
=&\f{\as}{2\pi}\left(\f{\pi}{\sin\pi\g}\right)^2
\cos(\pi\g)\f{2+3\g(1-\g)}{(3-2\g)(1-2\g)(1+2\g)}
\end{align}
so that we finally get, for massless quarks, in the small $\o$ limit
\begin{equation}
G_\o^{qA}(Q)\simeq N_f\int\f{d\g}{2\pi i}H_2(\g)f_\o^A(\g)e^{\g t}+
\text{low energy terms}
\end{equation}
which becomes, in the anomalous dimension regime,
\begin{equation}
G_\o^{qA}(Q)=\sum_{\bar{\g}}\bar{\g}H_2(\bar{\g})G_{\o\bar{\g}}^{gA}(Q).
\end{equation}

The saddle points $\bar{\g}(\as(t))$ are given here by all solutions of the 
equation (2.14), i.e.,
\begin{equation}
1=\f{\bar{\as}(t)}{\o}\left[\chi_0(\bar{\g}(\as))+\as\chi_1(
\bar{\g}(\as))\right]
\end{equation}
where we have explicitly shown the NL term of the BFKL eigenvalue.

While at leading level we only have the contribution of the BFKL anomalous 
dimension $\g_L$
\begin{equation}
\begin{cases}
1 & = \f{\bar{\as}}{\o}\chi_0(\g_L)\\
\g_L & = \f{\bar{\as}}{\o}+2\zeta(3)\left(\f{\bar{\as}}{\o}\right)^{4}+\dots
\end{cases}
\end{equation}
at NL level we expect saddle points located at the two eigenvalues of 
the anomalous dimension matrix $\g_{ab}$, i.e., 
\begin{equation}
\g_+\simeq\g_{gg}+\f{C_F}{C_A}\g_{qg},\quad\g_-\simeq \g_{qq}-\f{C_F}{C_A}
\g_{qg}=-\f{C_F}{C_A}\g_{qg}^{(1)}+\dots
\end{equation}
where we have used the fact that, at leading level 
$\g_{gq}\simeq\f{C_F}{C_A}\g_{gg}$, and that the NL $\g_{qg}$ starts at 
one loop level ($\g_{qg}^{(1)}$).

Here a subtle point arises. The NL hierarchy of the BFKL kernel really 
works only around the leading eigenvalue $\g_+$, but not around $\g_-$. 
In fact, subleading BFKL kernels of order $n\geq 1$ (the leading one 
corresponding to $n=0$) show multiple collinear singularities of order 
$\f{1}{\o}\left(\f{\as}{\g}\right)^{n+1}$. Evaluating the latter around 
$\g_L\simeq\bar{\as}/\o$ yields genuinely subleading terms of relative 
order $\o^n$. Instead, for small values of $\g\simeq \g_-\sim \as$ 
all subleading kernels become of the same order!

This means that, in order to recover the correct collinear behaviour 
of the partonic Green' s functions around $\g=\g_-$, we need to resum a whole 
series of collinear singular BFKL kernels. For instance, at one loop level,
the analogue of $f_\o^{ba}(\g)$ in Eq. (2.7) ($a,b=q,g$) 
is simply found from the DGLAP 
equations. Keeping $\as$ frozen for simplicity we obtain
\begin{equation}
f_\o^{ba}(\g)=
\f{1}{\g}\left(1-\f{\G}{\g}\right)^{-1}=
\f{1}{\g\left(1-\f{\g_+}{\g}\right)
\left(1-\f{\g_-}{\g}\right)}
\begin{pmatrix}
1-\f{\g_+}{\g} & \f{\g_{qg}}{\g} \\
\f{\g_{gq}}{\g} & 1
\end{pmatrix}
\end{equation}
which satisfies a BFKL-type equation of the form (2.6), with
\begin{equation}
k=\left(\g-\hat{\G}\right)^{-1},\quad\quad
K=kP_L,
\end{equation}
where $P_L$ and $\hat{\G}$ are defined by the expression
\begin{equation}
\G=\f{\bar{\as}}{\o}P_L+\hat{\G},
\end{equation}
which is the NL decomposition of the one-loop flavour singlet anomalous 
dimension matrix (Cfr. Eq. (1.3)).
It is then easy to check that the collinear resummation in the kernels of 
Eq. (2.32) is essential to provide the correct singular behaviour 
at $\g=\g_-$, thus cancelling a variety of singularities around $\g=0$ 
that finite order expansions have.

On the other hand, since we are working at NL level, the one-loop 
result is all we need to describe $\g_-$, which does not have enhanced higher 
order contributions. 
This means that Eqs. (2.32)-(2.34) are enough to find the correct 
behaviour around $\g_-$ if running coupling effects are introduced 
according to the renormalization group.

The output of this discussion is that, in Eqs. (2.18) and (2.26) we evaluate 
the saddle point $\g_+$ by the $\k$-factorization formulae given before, and 
we evaluate $\g_-$ by the known one-loop collinear behaviour, because 
higher order enhancement effects are not present. 
The resulting quark-sea and gluon densities are then given by
\begin{equation}
\begin{pmatrix} G^{qa}\\ G^{ga}\end{pmatrix}=
\begin{pmatrix} N_f\g_+ H_2(\g_+)\\ 1 \end{pmatrix}G_+^{ga}+
\begin{pmatrix} -\f{C_A}{C_F}\\ 1 \end{pmatrix} G_-^{gA},
\end{equation}
where
\begin{equation}
\begin{cases}
G_+^{gA}=\f{1}{\g_+\sqrt{-\chi^\prime(\g_+(\as(t)))}}\exp
\left[\int_0^t\g_+(\as(t))dt\right]K_{+}^A\\
G_-^{gA}=\exp\left[\int_0^t\g_-(\as(t))dt\right]K_{-}^A
\end{cases}
\end{equation}

The final result, Eq. (2.35) is consistent with 
the renormalization group with 
a properly defined anomalous dimension matrix $\G$. In particular, the 
$t$-dependent $2$-vectors of Eq. (2.35) are the right-eigenvectors of $\G$, 
so that we directly obtain, with simple algebra, the following 
conclusions:
\begin{enumerate}
\item The off-diagonal entry $\g_{qg}$ is given by \cite{10,15}
\begin{equation}
\g_{qg}=N_f\g_+^2H_2(\g_+)=N_f\g_L^2\left(\f{\bar{\as}}{\o}\right)
H_2\left(\g_L\left(\f{\bar{\as}}{\o}\right)\right)~~+~~\text{NNL}
\end{equation}
with $H_2(\g)$ given by Eq. (2.26).
\item The larger eigenvalue of $\G$, call it $\g_+^{t_0t}$, is not 
exactly the leading solution $\g_+$ of Eq. (2.29), but is given by
\begin{equation}
\g_+^{t_0t}=\g_+(\as(t))-\f{d}{dt}\log\left[\g_+
\sqrt{-\chi^\prime(\g_+)}\right],
\end{equation}
where we have included in the anomalous dimension the NL contribution 
due to the running coupling in the coefficient function $H(\as)$.
\item The lower eigenvalue 
$\g_-=-\f{C_F}{C_A}\g_{qg}^{(1)}$ is just the one loop result, 
so that also
\begin{equation}
\g_{qq}=\f{C_F}{C_A}\left(\g_{qg}-\g_{qg}^{(1)}\right)
\end{equation}
\end{enumerate}

Since the results (2.37) and (2.39) have been discussed elsewhere \cite{10,15}, 
we concentrate in the following on the new results on $\g_+$ of Eq. (2.29)
and (2.38), 
obtained by evaluating the NL eigenvalue $\chi_1$.
Note that the combination $G^q+\f{C_F}{C_A}G^g$ evolves according to 
$\g_+^{t_0t}$, given in Eq. (2.39) and that the real dynamical 
contribution in the latter 
is the one due to $\g_+$, which will be discussed in the following sections.

On the other hand, the contribution to (2.38) 
due to the logarithmic derivative of $H$ starts from 
three loops on, and is a peculiarity of the BFKL equation arising from
its $\k$-fluctuations, which become large when $\g_L$ approaches the 
saturating value $\g_L=1/2$\footnotemark.
\footnotetext{Although the $t$-dependent $\o$-singularity of the 
anomalous dimension may be at a slightly different position than 
$\o_{\mathbb{P}}(t)$ \cite{20,22}, it remains true that resummation 
effects are large when $\g_L\left(\f{\as(t)}{\o}\right)$ approaches 
$1/2$.}
This point was discussed elsewhere \cite{15} for frozen $\as$ and arises 
from the saddle-point features of Eq. (2.11) in the running $\as$ case.

To sum up, we have given in this section a definition of the gluon and quark
densities based on $\k$-factorization only, i.e., we have defined the 
so-called $Q_0$-scheme. We have also derived from the NL BFKL equation 
(2.1) the anomalous dimension representation (2.35) 
in the regime (2.13), leading to 
the expression (2.37)-(2.39) of the anomalous dimensions.

We have thus set up the framework for deriving resummation formulas from 
the properties of the kernel eigenvalue on the basis of the 
definition of $\g_+$ given in Eq. (2.29). 

\newpage
\resection{Real $q\bar{q}$ Emission Kernel}
The program of computing the  
NL kernel \cite{11} 
involves the lowest order computation of the  
2-particle $s$-channel discontinuity of the four gluon exchange amplitude, 
of the one loop correction to the corresponding 1-gluon intermediate state,
and of the purely virtual two loop corrections  
(Fig. 3). Each of these terms can be further divided
in sub-terms having different colour structure,  
according to their dependence on $C_A$, $C_F$, $N_f$.
In the following we will 
concentrate on the $N_f$-dependent part, which will be called the 
$q\bar{q}$ contribution.

Up to now all virtual corrections (to one or no-particle discontinuities) 
have been computed by Fadin and collaborators \cite{13,14}, while the 
evaluation of the real 
emission terms has not been completed. In fact, amplitudes \cite{8,11} 
and squared matrix elements  
for $q\bar{q}$ \cite{8} and gluonic \cite{12} contributions 
are known in the literature, but the  
phase-space integrations have not yet been performed.  

In the following sections 
we give the details of our recent computation \cite{16} 
of the $q\bar{q}$ emission 
kernel, of its eigenvalues, and of the corresponding anomalous dimension.

\resub{Massive $q\bar{q}$ production}

In order to regularize $s$-channel collinear singularities, we will
work with massive quarks (the relevant formulas in the massless limit and 
$4+2\ep$ dimensions will be given in the next section). As an 
intermediate step, we shall obtain the high energy coefficient function
for the $Q\bar{Q}$ production via gluon-gluon fusion.

We thus start by considering the total cross section for the hadroproduction
of a (heavy) quark-antiquark pair, which at high energies is dominated by 
the (Regge) gluon fusion process of Fig. 4. 
In the large center of mass energy limit the perturbative expansion of this 
cross section contains large terms of order $\left[\as(M^2)\log
(s/M^2)\right]^n$, 
which need to be resummed to all orders. This resummation is 
performed by using the $\k$-factorized expression for the cross section 
\cite{8}
\begin{equation}
M^2\ss_{12}=\int\f{dz_1}{z_1}\f{dz_2}{z_2}d^2\k d^2\k^\prime
\ssh(\k,\k^\prime,M^2,z_1z_2s)
\FF_1(z_1,\k)\FF_2(z_2,\k^\prime),
\end{equation}
in which the gluon densities $\FF_i$ satisfy the BFKL equation (2.1), and
$\ssh$ is the (lowest-order and gauge-invariant) off-shell  
continuation of the partonic cross section. 

In the language of $Q^2$ and $s$-moments defined in Sec. 2, we 
introduce the double Mellin transformed version of Eq. (3.1), i.e., 
\begin{align}
&M^2\ss\left(\f{M^2}{s}\right)=
\int_{\f{1}{2}-i\infty}^{\f{1}{2}+i\infty}\f{d\g_1d\g_2d\o}{-i(2\pi)^3
\g_1\g_2}\left(\f{M^2}{Q_0^2}\right)^{\g_1+\g_2}\FF_{1\o}(\g_1)
\FF_{2\o}(\g_2)H_\o(\g_1,\g_2) \\
\intertext{where}
&\FF_{i\o}(\g)=\int^1_0dzz^{\o-1}\int_0^\infty\f{d^2\k}{\pi\k^2}
\left(\f{\k^2}{Q_0^2}\right)^\g\FF_i(z,\k)\\
&H_\o(\g_1,\g_2)=\int_0^\infty\f{d^2\k}{\k^2}\f{d^2\k^\prime}{\k^{\prime 2}}
\left(\f{\k^2}{M^2}\right)^{\g_1}\left(\f{\k^{\prime 2}}{M^2}\right)^{\g_2}
\int_0^\infty\f{ds}{s}\left(\f{M^2}{s}\right)^\o \ssh(\k,\k^\prime,M^2,z_1z_2s)
\end{align}
For large enough $M^2$ the $\g$-integrals are dominated by the BFKL anomalous 
dimensions (2.31), so that the process-dependent high-energy resummation 
effects are embodied in the coefficient function $H_\o(\g_1,\g_2)$, 
through the $\o$-dependence of $\g_1$ and $\g_2$. Part of this 
section is devoted to the evaluation of this coefficient function.

Let us start by defining the kinematics of the process under consideration
(Fig. 4).
First of all we introduce a Sudakov parametrization of the exchanged 
gluons' momenta
\begin{equation}
k^\mu\simeq z_1p_1^\mu+\k^\mu,\quad\quad
k^{\prime\mu}\simeq -z_2p_2^\mu-\k^{\prime\mu},
\end{equation}
and of the momentum transfer
\begin{equation}
\Delta^\mu=z_1x_1p_1^\mu-z_2x_2p_2^\mu+\dt^\mu,
\end{equation}
where $p_1$, $p_2$ denote the (light-like) momenta of the incoming 
hadrons. Expressing the phase-space 
in terms of these variables, the moments in Eq. (3.4) take the form
\begin{align}
&\bks\f{\as}{\pi}H_\o(\g_1,\g_2)=\nn\\
&\bks \int\f{d\nu}{\nu^{2-\o}}\f{d^2\k}{\pi\k^2}
\f{d^2\k^\prime}{\pi\k^{\prime 2}}dx_1dx_2d^2\dt
\delta(x_2(1-x_1)\nu-(\k-\dt)^2-M^2)\times\nn\\
&\times\delta(x_1(1-x_2)\nu-(-\k^\prime+\dt)^2-M^2)
\left(\f{\k^2}{M^2}\right)^{\g_1}
\left(\f{\k^{\prime 2}}{M^2}\right)^{\g_2}
A(\k,\k^\prime,\dt,x_1,x_2,\nu,M^2).
\end{align}
Here $A$ denotes the (off-shell) squared matrix element for the
(Regge) gluon fusion channel of the $Q\bar{Q}$ production process, 
and we have defined
\begin{equation}
\nu=z_1z_2(2p_1\cdot p_2)=z_1z_2s.
\end{equation}

In order to understand the colour and singularity structure of 
the $q\bar{q}$ kernel, in view of the following phase-space integrations, 
we find it convenient to rewrite the squared 
matrix element of Ref. \cite{8} in 
the following form:
\begin{align}
\f{A}{\nu^2}&=\alpha_s^2(D_1+D_2+D_3)\\
\intertext{where}
D_1&=(2C_F-C_A)\left[\f{-1}{(M^2-\th)(M^2-\uh)}+
\f{(B+C)^2}{\k^2\k^{\prime 2}}\right],\nn\tag{3.10a}\\
D_2&=C_A\left[\f{1}{\sh}\left(\f{1}{M^2-\uh}-
\f{1}{M^2-\th}\right)(1-x_1-x_2)-
\f{B^2+C^2}{2\k^2\k^{\prime 2}}+\right.\nn\\
&\left.\quad\quad\quad\quad+\f{2(B-C)}{\k^2\k^{\prime 2}\sh}\left(
\k^2(1-x_2)+\k^{\prime 2}(1-x_1)+\k\cdot\k^\prime\right)\right],\nn\tag{3.10b}\\
D_3&=C_A\left[\f{2}{\nu\sh}-\f{2}{\k^2\k^{\prime 2}}
\f{\left(\k^2(1-x_2)+\k^{\prime 2}(1-x_1)
+\k\cdot\k^\prime\right)^2}{\sh^2}\right],\nn\tag{3.10c}\\
\intertext{and we have introduced the notation}
B=&\f{1}{2}-\f{(1-x_1)(1-x_2)\nu}{M^2-\th}+\f{\nu(1-x_1-x_2)}{2\sh}+
\f{\dt\cdot(\k+\k^\prime)}{\sh},\nn\tag{3.11a}\\
C=&\f{1}{2}-\f{x_1x_2\nu}{M^2-\uh}-\f{\nu(1-x_1-x_2)}{2\sh}-
\f{\dt\cdot(\k+\k^\prime)}{\sh}\nn\tag{3.11b},
\end{align}
\addtocounter{equation}{2}
and the Mandelstam variables of the hard sub-process
\begin{equation}
\sh=(k_1+k_2)^2,\quad\th=\Delta^2\quad\uh=(k_1-k_2-\Delta)^2.
\end{equation}

We have decomposed the matrix element (3.9) in the three terms (3.10) in such 
a way that the integrals over $\nu$ in Eq. (3.7) are separately 
convergent. Furthermore, all the $s$-channel collinear singularities are 
contained in the term $D_3$ in Eq. (3.10c), since the various poles of the 
type $1/\sh$ and $1/\sh^2$ in (3.10b) cancel each other. This implies 
that we will be allowed to compute the contribution of $D_1$ and $D_2$
to the kernel directly in the massless limit and four space dimensions, 
and only the term $D_3$, with its simple structure, will need a 
regularization.

As a final remark, note that the first term ($\sim (2C_F-C_A)$)
is suppressed by a colour factor of $1/C_A^2$ with respect to the others. 
We shall see in the 
following that this suppression factor is relevant in order to assess the 
magnitude of NL $q\bar{q}$ contributions to the kernel eigenvalue.

The algebra involved in the evaluation of the phase-space integrals 
in Eq. (3.7) is cumbersome. So we give here only the final result in the 
$\o=0$ limit, and we report the details of the phase-space integrations
in Appendix A. Following the decomposition (3.9) of the matrix element, we
split the moment in Eq. (3.4) in three parts $H^{(1)}$, $H^{(2)}$ and
$H^{(3)}$, as follows
\begin{align}
&\bks H_0(\g_1,\g_2)=H^{(1)}(\g_1,\g_2)+H^{(2)}(\g_1,\g_2)+
H^{(3)}(\g_1,\g_2)\\
\intertext{where}
&\bks H^{(1)}=(2C_F-C_A)H^{ab}(\g_1,\g_2)\nn\tag{3.14a}\\
&\bks H^{(2)}=\f{\as}{\pi}C_AB(\g_1,1-\g_1)B(\g_2,1-\g_2)
\left(\f{\G(2-\g_1-\g_2)}{\G(4-2\g_1-2\g_2)}-
\f{\G(3-\g_1-\g_2)}{\G(6-2\g_1-2\g_2)}\right)\nn\tag{3.14b}\\
&\bks H^{(3)}=\f{\as}{\pi}C_A\f{B(\g_1,1-\g_1-\g_2)B(\g_2,1-\g_1-\g_2)
B(2-\g_1-\g_2,2-\g_1-\g_2)}{2(1-\g_1-\g_2)}\times\nn\\
&\bks\times\left(1-\f{1}{5-2\g_1-2\g_2}\right)\tag{3.14c}\nn
\end{align}
\addtocounter{equation}{1}
and $H^{ab}(\g_1,\g_2)$ is the abelian contribution (2.26).

The result (3.13)-(3.14) provides, 
in $\g$-space, the leading order coefficient for the production 
of a (heavy) quark-antiquark pair via gluon-gluon fusion, relevant for 
$Q\bar{Q}$ production in hadron-hadron scattering, and completes the 
results of Ref. \cite{8} where the $C_F$ dependent part was computed.

The above coefficient (in the massless limit) is related to the NL
eigenvalue,
representing the action of the kernel on the leading order powerlike 
eigenfunctions $\sim(\k^2)^{\g-1}$. The massless limit, obtained as the 
residue of the pole at $\g_1+\g_2=1$ (Cfr. Eqs. (2.25) and (2.26)), 
is singular. 
In fact, due to the $s$-channel collinear singularity of $D_3$, 
the term $H^{(3)}$ in Eq. (3.14c) has triple and double poles, which must be 
cancelled by the virtual terms.

\resub{Massless quark limit}

The triple pole singularity of $H^{(3)}$ in Eq. (3.14c) 
is due to the singular behaviour 
of the $gg\ra q\bar{q}$ cross section of Eq. (3.4)
in the region $M^2\ll(\k-\k^\prime)^2\equiv\q^2\ll\k^2\simeq\k^{\prime 2}$. 
In fact, after phase space integration, the partonic cross section 
approaches the singular limit 
\begin{align}
&\bks\ssh_{\o=0}(\k,\k^\prime,M^2)=\nn\\
&\bks\bar{\as}\f{\as}{\pi}\int_0^1
dx_1\f{x_1(1-x_1)}{\q^2}\log\left(1+\f{x_1(1-x_1)\q^2}{M^2}\right)
\simeq\bar{\as}\f{\as}{6\pi}\left[\log\f{\q^2}{M^2}-\f{5}{3}\right]\f{1}{\q^2}
\end{align}
which is 
responsible for the triple pole, of the $\k^2$-moments in Eq. (3.13).

As remarked before, this singularity comes from the last 
term $D_3$ 
in Eq. (3.10), while the $s$-channel singularities 
of $D_2$ cancel each other after phase-space integration.

It is interesting to notice that such singular behaviour 
is responsible for a sizeable enhancement of the $Q\bar{Q}$
hadroproduction cross section at high energies when the two anomalous dimension 
$\g_1$ and $\g_2$ approach the saturating value of $1/2$. On this basis 
we expect relevant effects due to the resummation of higher orders in
the b-quark cross-section at Tevatron, 
which will be the subject of further investigations.

From the point of view of the $q\bar{q}$ kernel instead, we have still to 
include the virtual contributions, which are singular also.
Having identified the singular real emission part 
with the $H^{(3)}$ term in Eq. 
(3.14c), we subtract it off and we perform the massless limit on the 
finite part, leaving to the next section the combination of $H^{(3)}$ 
with the virtual radiative corrections, to yield an overall finite result.

In analogy with Eq. (2.25) we then 
define the regular $q\bar{q}$ contribution to 
the kernel eigenvalue, as follows
\begin{align}
\chi_{reg}^{q\bar{q}}&=\underset{\g_1\ra 1-\g}{\lim}
(1-\g-\g_1)\left[H^{(1)}(\g,\g_1)+H^{(2)}(\g,\g_1)\right]=\nn\\
&=\f{C_A\as}{6\pi}\left(\f{\pi}{\sin\pi\g}\right)^2
\left(1-\f{3(2C_F-C_A)}{C_A}\cos(\pi\g)
\f{1+\f{3}{2}\g(1-\g)}{(3-2\g)(1-2\g)(1+2\g)}\right),
\end{align}
which is obtained directly from Eqs. (3.14a) and (3.14b).

For completeness we report the kernel (3.16) in $\k$-space also.
By inverting Eq. (2.4), as explained in Appendix B, 
we are able to express the azimuthal averaged 
real emission regular $q\bar{q}$ kernel as follows,
\begin{align}
K_{reg}^{q\bar{q}}&=\f{1}{\k^2}\int_{\f{1}{2}-i\infty}^{\f{1}{2}+i\infty}
\f{d\g}{2\pi i}\chi_{reg}^{q\bar{q}}(\g)\left(\f{\k^2}{\k^{\prime 2}}
\right)^{-\g}=\nn\\
&=\f{2C_F-C_A}{C_A}\f{N_f\as}{64\pi\k_>^2}
\left[\log\f{1}{\rho}\left(\f{1}{\rho}-1\right)+2\left(\f{1}{\rho}+1\right)+
\right.\nn\\
&\left.+2\left(\LL_2(\rho)+\log\f{1}{\rho}\LL_1(\rho)\right)
\left(22-\rho-\f{1}{\rho}\right)\right]+\nn\\
&+\f{C_A\as}{6\pi}\f{1}{\k^2-\k^{\prime 2}}\log\f{\k^2}{\k^{\prime 2}}.
\end{align}
where
\begin{align}
\LL_1(\rho)=&\f{1}{\sqrt{\rho}}\arctan(\sqrt{\rho}),\\
\LL_2(\rho)=&\f{1}{2i\sqrt{\rho}}\left(Li_2(i\sqrt{\rho})-
Li_2(-i\sqrt{\rho})\right),\\
\rho=&\f{k_<^2}{k_>^2},\quad\quad
k_<=\min(|\k|,|\k^\prime|),\quad k_>=\max(|\k|,|\k^\prime|),
\end{align}
and $Li_2$ denotes the Euler Dilogarithm.

\newpage
\resection{Complete $q\bar{q}$ Kernel and Anomalous Dimension}
\resub{Combining with virtual terms}

Let us now consider the contribution to the $q\bar{q}$ kernel of the singular 
part in Eq. (3.10c), whose $\g$-moments yield the term $H^{(3)}$
in Eq. (3.14c) with the triple pole.
The singular behaviour for $M^2$ and $\q^2\ra 0$ of the cross section (3.15) 
is cancelled in the total NL kernel by the corresponding singularities in the 
virtual terms.

In order to cancel the $M^2=0$ singularity we need the quark loop 
contribution to the one-gluon 
$s$-channel discontinuity, which is simply obtained 
from the NL reggeon-reggeon-gluon vertex computed in Ref. \cite{13}. 
By squaring the corresponding amplitude, 
summing over the polarizations of the 
emitted gluon and averaging over the azimuth, we get
\begin{equation}
\f{\as N_f}{6\pi}
\left(\f{1}{\q^2}\log\f{M^2}{\mu^2}-\f{1}{\k^2-\k^{\prime 2}}
\log\f{\k^2}{\k^{\prime 2}}\right),
\end{equation}
in which $\mu^2$ is the renormalization scale.
By combining with the singular $q\bar{q}$ emission part (3.14) we get the 
$M^2$-independent expression
\begin{equation}
\f{\as N_f}{6\pi}\left[\left(\log\f{\q^2}{\mu^2}-\f{5}{3}\right)\f{1}{\q^2}
-\f{1}{\k^2-\k^{\prime 2}}\log\f{\k^2}{\k^{\prime 2}}\right].
\end{equation}
which is still singular for $\q^2=0$. 

The same computation can be performed in the massless quark limit and 
$4+2\ep$ dimensions, to yield the $\ep$-dependent kernel 
(see Appendix C)
\begin{equation}
\f{\as N_f}{6\pi}\left[\f{1}{\ep}\f{\G^2(1+\ep)}{\G(1+2\ep)}
\left(\f{\q^2}{\mu^2}\right)^\ep\f{1-\f{5}{3}\ep+\f{28}{9}\ep^2+0(\ep^3)}
{\q^2}-\f{1}{\ep\q^2}-\f{1}{\k^2-\k^{\prime 2}}\log\f{\k^2}{\k^{\prime 2}}
\right].
\end{equation}

In order to cancel the $\q^2=0$ singularity we need to combine the above 
results with the purely virtual terms, which were computed in dimensional 
regularization in Ref. \cite{14}. This can be done by means of Eq. (4.3), 
and after the cancellation of the singularities at $\ep=0$, the 
NL $q\bar{q}$ contribution to the kernel can be expressed 
directly in 4 dimensions as follows (see Appendix C):
\begin{align}
K_0+K_1^{q\bar{q}}=&\left(1+\f{\as N_f}{6\pi}\log\f{\k^2}{\mu^2}\right)
K_0(\k,\k^\prime)+\nn\\
&+\f{\as N_f}{6\pi}\left[\left.\left(\log\f{\q^2}{\k^2}-\f{5}{3}\right)
\f{1}{\q^2}\right|_R-
\f{1}{\k^2-\k^{\prime 2}}\log\f{\k^2}{\k^{\prime 2}}\right]+
K_{reg}^{q\bar{q}}(\k,\k^\prime),
\end{align}
where $K^{q\bar{q}}_{reg}$ is the regular part of the real emission kernel 
given in Eq. (3.17), and we have factorized the $q\bar{q}$ contribution to 
the running of $\as$ at the scale $\k^2$ in agreement with the choice of 
Sec. 2. 

We remark that such a choice of the scale of $\as$ is not strictly 
consequence of 
our NL calculation, since it fixes only the coefficient of the 
$\log(1/\mu^2)$ term, while the upper scale of the logarithm can be changed 
by changing the scale invariant part of the kernel. However this choice 
allows the clear discussion of factorization properties of the BFKL 
equation provided in Sec. 2 (Eqs. (2.35) and (2.36)), 
and yields the NL coefficient (2.19), 
induced by the running coupling, which differs from unity only 
from three loops on.  

Finally, we are now able to compute the finite eigenvalue of the scale 
invariant kernel $K_1$ in Eq. (4.4). By using standard identities of type
\begin{equation}
\f{1}{\k^2-\k^{\prime2}}\log\f{\k^2}{\k^{\prime2}}=
\int_0^1\f{dy}{y\k^2+(1-y)\k^{\prime2}}\nn\tag{4.5a}
\end{equation}
and
\begin{align}
\int_0^\infty\f{d^2\k}{\pi\k^2}\left(\f{\k^2}{M^2}\right)^\g
\f{1}{(M^2+\xi\k^2)^n}=(M^2)^{-n}\xi^{-\g}B(\g,n-\g),\nn\tag{4.5b}
\end{align}
\addtocounter{equation}{1}
we obtain the expression
\begin{align}
\chi_1^{q\bar{q}}=&\f{\as N_f}{6\pi}\left[\f{1}{2}(\chi_0^\prime(\g)+
\chi_0^2(\g))-\f{5}{3}\chi_0(\g)-\right.\nn\\
&\left.-\f{1}{N_C^2}\left(\f{\pi}{\sin\pi\g}\right)^2
3\cos(\pi\g)\f{1+\f{3}{2}\g(1-\g)}{(3-2\g)(1-2\g)(1+2\g)}\right]
\end{align}
which is plotted in Fig. 5 as a function of $\g$.

Note that the expression (4.6) is symmetrical for $\g\ra 1-\g$, 
apart from the $\chi^\prime_0(\g)$ term, which is due to our choice of 
$\k^2$ as scale of $\as$, instead of a more symmetrical combination of 
$\k$ and $\k^\prime$.

\resub{Eigenvalue properties and anomalous dimension}

From the plot of Fig. 5 it is evident that a cancellation between terms with 
different colour structure is at work. In fact, the last 
term in the eigenvalue (4.6) is essentially the abelian one of Eq. (2.26), 
except that it is multiplied by the 
small colour factor $1/N_C^2$. This means 
that the non planar diagrams are the ones responsible for the 
large resummation effects,
giving large contributions at high values of both $\k^2$ and 
$\k^{\prime2}$.
In fact, they are suppressed in the large $N_C$ limit when coupled to 
gluons in $K_1$, but are not when coupled to photons in $H_2$ (Eq. (2.26)).

From Eq. (4.6) the part of the shift of the pomeron intercept which is 
proportional to $N_f$ can be estimated. 
For models with a continuum spectrum \cite{20} it is simply 
obtained by setting $\g=1/2$, to yield
\begin{align}
&\Delta\op=\op\f{\as N_f}{6\pi}\left[2\log 2-\f{5}{3}-\f{1}{N_C^2}
\f{33}{32}\f{\pi^3}{8\log 2}\right],\\=
&\f{\Delta\op}{\op}\simeq-0.05N_f\as,
\end{align}
which is just a few percent for reasonable values of $N_f$ and $\as$. 
Once again, the potentially large term $\sim\pi^3$ is suppressed
by a colour factor.

In order to compute the NL correction to the anomalous dimensions, we 
introduce the 'renormalized' coupling
\begin{equation}
\bar{\as}\ra\tilde{\as}=\bar{\as}\left(1+\f{\Delta\op}{\op}\right).
\end{equation}
in the defining equation (2.29), so that it takes the form
\begin{equation}
1=\f{\tilde{\as}}{\o}\left[\chi_0(\g_+)+\as\left(\chi_1(\g_+)-\chi_1(1/2)
\right)\right]
\end{equation}
and then we expand $\g_+$ in the $\chi_0$-term around 
$\g_+=\g^L(\tilde{\as})$, to get
\begin{align}
&\left[\g_+-\g_L\right]^{q\bar{q}}=\as\f{
\chi_1^{q\bar{q}}(\g_L)-\chi_1^{\q\bar{q}}(1/2)}{-\chi_o^\prime(\g_L)},\\
&1=\tilde{\as}\chi_0(\g_L).
\end{align}
In this way the perturbative expansion in powers of 
$\g_L(\tilde{\as})$ is free of spurious singularities for $\g_L=1/2$, which 
would signal precisely a renormalization of the Pomeron intercept.

The NL correction (4.11) to the leading eigenvalue of the anomalous dimension 
matrix $\g_+$ is plotted in Fig. 6 as a function of $\g_L(\tilde{\as})$.
This correction is small compared to the leading 
anomalous dimension $\g_L$, apart from the region of very small $\g_L$ 
corresponding to $\o\geq\as$, where it is needed in order to achieve 
consistency with fixed order renormalization group equation. 
Therefore, higher order effects are negligible.

A particular comment is needed on how fixed order perturbation theory is 
recovered in this formalism.
In fact, since $\g_L=\bar{\as}/\o+O(\bar{\as}^4)$, the small $\g$ 
behaviour of $\chi_1(\g)$ in 
Eq. (4.11) determines directly the one and two loop 
anomalous dimensions:
\begin{align}
\g_+-\g_L\simeq\f{\as N_f}{6\pi}\left[-\f{1}{N_C^2}-\f{5}{3}
\left(1+\f{13}{10N_C^2}\right)\f{\bar{\as}}{\o}+....\quad\right]\\
\intertext{and from Eq. (2.31) we obtain}
\g_{gg}\simeq-\f{\as N_f}{6\pi}\left(1+\f{23}{6}\f{\bar{\as}}{\o}+
....\quad\right)
\end{align}
which agrees with the two loop computation in the DIS scheme \cite{26}.

More in general, we can say that if 
\begin{equation}
\g_+^{(1)}=\f{\bar{\as}}{\o}+A_1\as,\quad\quad\g_+^{(2)}=A_2\as
\f{\bar{\as}}{\o}
\end{equation}
is the NL expression at two loop order, 
the eigenvalue $\chi_1$ should have the
structure
\begin{equation}
\chi_1(\g)\simeq\f{A_1}{\g^2}+\f{A_2}{\g}+(\g\ra 1-\g)+
\text{running coupling terms}
\end{equation}
just on the basis of the $\g=0$ singularities needed to recover (4.14)
and of the basic $\k\ra\k^\prime$ symmetry of the BFKL kernel - apart 
from the asymmetry induced by  factorizing out $\as(\k^2)$ in Eq. (4.4).

The singularities at $\g=0$ occurring in Eq. (4.16) are just the collinear
ones coming from transverse momenta $|\k|\gg\Delta\gg|\k^\prime|$, 
while the ones at $\g=1$ come from the disordered region 
($|\k^\prime|\gg\Delta\gg|\k|$), that we have already 
emphasized \cite{25} as a possible source of large higher order effects.
The third possible source of collinear singularities - the one in the 
$s$-channel - turns out to combine with virtual corrections to give rise
to running coupling effects added in Eq. (4.16).

The structure of Eqs. (4.15) and (4.16) is interesting for various 
reasons. Firstly, it appears that the NL terms arise from various collinear
singular kernels, as already emphasized in Eqs. (2.33)-(2.34) at  
one-loop level. This is because the two channel collinear factorization 
is not disentangled yet in the BFKL approach (Cfr. Sec. 2.2).

Secondly, we can follow the suggestion of Ref. \cite{25}, and 
use the expression (4.16), including the $(1-\g)$-terms, as a crude 
estimate of $\chi_1$ for the gluonic contribution, where a complete 
calculation is not yet available. In that case we should set
\begin{equation}
A_1=-\f{\as N_f}{6\pi}\f{1}{N_C^2}-\f{11N_C\as}{12\pi},\quad\quad
A_2=-\f{\as N_f}{6\pi}\left(\f{5}{3}+\f{13}{6N_C^2}\right)
\end{equation}
with the rough estimate
\begin{equation}
\chi_{1,coll}=\chi_{1,coll}^{(q\bar{q})}-\f{11N_C\as}{12\pi}
\left(\f{1}{\g^2}+\f{1}{(1-\g)^2}\right)+....
\end{equation}

The last term in Eq. (4.18) is large, and - if extended to large $\g$ 
values - would provide a negative and sizeable shift of the Pomeron 
intercept and important resummation effects, unlike what happens for the
$q\bar{q}$ contribution.

Although we do not think the estimate (4.18) is reliable when 
approaching $\g=1/2$, it provides an indication of the existence of 
unsuppressed planar contributions in the gluonic part. 
The complete calculation is therefore a rather important goal to achieve.

\newpage
\resection{Discussion}

We have studied in this paper the anomalous dimension 
representation for the NL BFKL equation and the ensuing resummation
formulas, by analysing in detail the $q\bar{q}$ contributions
to the larger anomalous dimension eigenvalue.

Our first result (Sec. 2) is that the R.G. representation with running 
coupling and resummed anomalous dimensions is valid at NL level in the 
regime
\begin{equation}
t=\log\f{Q^2}{\Lambda^2}\gg 1, \bar{b}\omega t >  \chi\left(\f{1}{2}\right),
\end{equation}
that is, if the variable $\alpha_s(t)\log\f{1}{x}$ is not too large.
If the condition (5.1) is coupled with the saddle point estimate of 
important $\omega$ values we end up with the limitation
\begin{equation}
\bar{\alpha}_S(t)\log\f{1}{x} < \f{1}{\chi\left(\f{1}{2}\right)}
\left(\log{\f{t}{t_0}}\right)t ,
\end{equation}
which provides a parabola-like boundary in the $\log\f{1}{x}$,
$\log\f{Q^2}{\Lambda^2}$ plane. 

Conditions of type (5.1) have 
been already noticed before \cite{22} 
as singularities of the anomalous dimension and those of
the type (5.2) have been known for a while \cite{21} to be relevant, 
with different numerical factors,
for the occurrence of higher twist unitarization effects [27-29].
Here we just emphasize that, within the regions
(5.1) and (5.2) there is a well defined resummation 
of anomalous dimensions, provided here, which is able to 
describe the QCD evolution in agreement with the NL BFKL equation (Fig 7).

The renormalization group description in the regime (5.1) holds
independently of the detailed properties of the hard Pomeron, i.e.
of the leading singularity in the $\omega$-plane which
dominates the small-$x$, fixed $Q^2$ behaviour. 
The latter occurs in the coefficient function, and is dependent 
on the soft region behaviour of the running coupling, in 
particular on its magnitude and shape around $\k^2=\Lambda^2$.
Only if the scale ${Q_0}^2$ is large enough, the Pomeron decouples, 
and the coefficient function takes a perturbative form, 
provided the rapidity is not so large to allow 
diffusion from ${Q_0}^2$ to $\Lambda^2$.

The simplest way to summarize the above results is to write the 
resummed R.G. representation of the DIS structure functions $F_2$ and $F_L$. 
By using formulas (2.35) 
for the parton densities in the $Q_0$-scheme we obtain
\begin{equation}
\begin{pmatrix} F_2^A(Q^2)\\ F_L^A(Q^2)\end{pmatrix}=
\begin{pmatrix} \g_+H_2(\g_+)\\ \g_+H_L(\g_+) \end{pmatrix}
G_+^A+
\begin{pmatrix} -\f{C_A}{C_F} \\ \f{\as}{3\pi} \end{pmatrix} G_-^A
\end{equation}
where $H_2$ and $H_L$ are given by Eq. (2.26) and by the relation \cite{10} 
\begin{equation}
H_L(\g)=\f{\g(1-\g)}{1+\f{3}{2}\g(1-\g)}H_2(\g)
\end{equation}
respectively, and ${G_\pm}^A$ are the R.G. expressions of Eq. (2.36).
In particular, 
\begin{equation}
G_+^A=\f{1}{\g_+\sqrt{-\chi^\prime(\g_+)}}\exp\left[\int_0^t
\g_+(\as(t))dt\right]
\end{equation}
contains the perturbative coefficient of Eq. (2.19), which provides an 
additional contribution to the effective anomalous dimension of Eq. 
(2.38).

Note that, once a complete NL computation of $\g_+$ will be available,
all the relevant coefficient and anomalous dimensions in 
Eq. (5.3) will be known in the $Q_0$-scheme, so as to 
provide a factorization scheme independent expression
for the measurable structure functions.

In fact, relating Eqs. (5.3) and (2.35) is equivalent to computing the 
quark and gluon coefficient functions in the $Q_0$-scheme to all leading 
orders. By simple algebra, and neglecting subleading contributions, we get
\begin{align}
&\bks C_q^2=1, &C_g^2=0;\\
&\bks C_q^L=\f{C_FN_f}{C_A}\left[\g_+H_L(\g_+)-\f{\as}{3\pi}\right], 
&C_g^L=\g_+H_L(\g_+).
\end{align}

In this paper we have further provided a computation of the 
$q\bar{q}$-contribution to the BFKL kernel, proving 
the running coupling factorization, and we have evaluated
the corresponding NL resummation in the anomalous dimension eigenvalue
$\g_+$.

We find that higher order effects for the $N_f$ dependent part are small in 
$\g_+$, while they are not small in the coefficients $H_2$ and $H_L$.
This difference is due to the non-planar nature of the diagrams which yield 
the most important large $\k$ contributions. They are suppressed by a 
colour factor when coupling to gluons, while they are not when coupling to 
photons.

We have also found that running coupling effects, although expected, are 
particularly important. First of all, they affect the coefficient functions
by factors of type $H$ in Eqs. (2.19) and (5.5) 
due to the fluctuations of the anomalous dimension variable 
$\g$, which are large when $\g$ approaches the saturation value 
$\g=1/2$. Ultimately, this is the reason for the breaking of the 
R. G. representation itself when approaching the critical value 
(5.1), as also noticed in a recent paper \cite{30}.

Furthermore, the running of $\as(t)\simeq 1/bt$ at large 
$t$ values, emphasizes the diffusion towards small values 
of $\k^2$, and the need of smoothing out the effective coupling 
around the Landau pole. This in turn clarifies the fact that the bare 
hard Pomeron singularity is actually dependent on soft physics, even if 
small-$x$ scaling violations are not.

It thus appears that, so far, apart from running coupling effects,
large higher order contributions only occur in the $\g_{qg}$ entry of the 
anomalous dimension matrix, which was the basis for an early explanation of
the HERA data \cite{4}. Note, however, that the absence of higher 
order effects in $\g_+$ may be a feature of the $N_f$-dependent
part only, which is mostly nonplanar. 
For the gluonic part, planar diagrams could also contribute, as it happens 
in the crude collinear estimate of Eq. (4.18). Such possibly large 
effects renew the present interest in a complete computation, 
which is hopefully to be obtained soon.

\begin{center}
{\bf Acknowledgements}
\end{center}

We wish to thank Jochen Bartels, 
Stefano Catani, Yuri Dokshitzer, Jan Kwiecinski, 
Al Mueller and Bryan Webber for interesting discussions on the topics 
presented here. This work is supported in part by MURST, Italy, and by 
the E. C. contract CHRX-CT96-0357.

\renewcommand{\theequation}{\Alph{section}.\arabic{equation}}
\newpage
\appsectio{Appendix A: Kinematics and Integrals}

We give here some details of the computation of the $H_0$ function (3.13).

Starting from the Sudakov parametrization (3.5)-(3.6) of the relevant 
momenta, we write the two-body phase space of the produced $q\bar{q}$ pair 
as follows
\begin{align*}
\bks\bks d\Phi(1,2)&=\f{\nu}{8\pi^2}dx_1 dx_2 d^2\dt
\delta(x_2(1-x_1)\nu-(\k-\dt)^2-M^2)\times\nn\\
&\times\delta(x_1(1-x_2)\nu-(-\k^\prime+\dt)^2-M^2)=\tag{A.1a}\\
&=\f{1}{8\pi}\f{dx_1}{x_1(1-x_1)}\f{d^2\tilde{\dt}}{2\pi}
\delta\left(\nu-\f{\tilde{\dt}^2+M^2}{x_1(1-x_1)}-\q^2\right)=\tag{A.1b}\\
&=\f{1}{8\pi}\f{dx_2}{x_2(1-x_2)}\f{d^2\hat{\dt}}{2\pi}
\delta\left(\nu-\f{\hat{\dt}^2+M^2}{x_2(1-x_2)}-\q^2\right),\tag{A.1c}
\end{align*}
where
\begin{align*}
&\tilde{\dt}=\dt-\k x_1-\k^\prime(1-x_1)\tag{A.2a},\\
&\hat{\dt}=\dt-\k^\prime x_2-\k(1-x_2).\tag{A.2b}.
\end{align*}
\addtocounter{equation}{2}
So, since the hard cross section is related to the matrix element by
\begin{equation}
\ssh(\k,\k^\prime,M^2,\nu)=\f{1}{2\nu}
\int d\Phi(1,2)A(\k,\k^\prime,\dt,x_1,x_2,M^2),
\end{equation}
we get, by (A.1a), the expression (3.7).

Since both parametrizations of the phase-space (A.1b) and (A.1c) are 
needed in the computation of the integrals, for each of them we give 
the expression of the relevant scalar quantities. 
Starting with the former, we have
\begin{align}
\sh=&\f{\tilde{\dt}^2}{x_1(1-x_1)},\\
M^2-\th=&(1-x_1)(\sh+\k^2)+x_1\k^{\prime 2}-2\tilde{\dt}\cdot\k^\prime,\\
M^2-\uh=&x_1(\sh+\k)^2+(1-x_1)\k^{\prime 2}-2\tilde{\dt}\cdot\k^\prime,\\
x_2=&\f{\left[(1-x_1)\q-\tilde{\dt}\right]^2+M^2}{(1-x_1)\nu},
\end{align}
for the latter
\begin{align}
\sh=&\f{\hat{\dt}^2+M^2}{x_2(1-x_2)},\\
M^2-\th=&(1-x_2)(\sh+\k^{\prime 2})+x_2\k^2-2\hat{\dt}\cdot\k,\\
M^2-\uh=&x_2(\sh+\k^{\prime 2})+(1-x_2)\k^2+2\hat{\dt}\cdot\k,\\
x_1=&\f{\left[(1-x_2)\q+\hat{\dt}\right]^2+M^2}{(1-x_2)\nu},
\end{align}
and for both of them
\begin{align}
\nu=&\sh+\q^2,\\
\q=&\k-\k^\prime.
\end{align}

In order to compute $H_0$, we find it convenient to
choose, for each term in the matrix element (3.9), the most suitable 
phase space
parametrization of type (A.1b) or (A.1c). The choice is done by 
requiring that
the denominator of each term in the matrix element (3.9) contains 
at most two of the invariants given before, and that at most one 
non-trivial azimuthal integration is needed. It is straightforward to 
check that at least one of our parametrizations satisfies these 
requirements for every term in Eqs. (3.9) and (3.10). 

With these premises we first use the $\delta$-function in (A.1b) or 
(A.1c) to eliminate the variable $\nu$, then, with the help of a 
Feynman parametrization of the denominators, we perform the integration over 
the transverse components of the momentum transfer $\dt$. In order to 
compute the $H_0$ function, the best way is to evaluate the 
$\k^2$-moments at this stage, since all the integrals needed can be 
reduced to the forms
\begin{align}
\int_0^\infty\f{d^2\k}{\pi\k^2}\left(\f{\k^2}{M^2}\right)^\g
\f{1}{(M^2+\xi\k^2)^n}=(M^2)^{-n}\xi^{-\g}B(\g,n-\g)
\end{align}
\begin{align}
&\int_0^\infty\f{d^2\k_1}{\pi \k_1^2}\f{d^2\k_2}{\pi\k_2^2}
\left(\f{\k_1^2}{M^2}\right)^{\g_1}
\left(\f{\k_2^2}{M^2}\right)^{\g_2}
\f{1}{(M^2+\xi_1\k_1^2+\xi_2\k_2^2)^n}=\nn\\
&=(M^2)^{-n}\xi_1^{-\g_1}\xi_2^{-\g_2}
\f{\G(\g_1)\G(\g_2)\G(n-\g_1-\g_2)}{\G(n)}
\end{align}
\begin{align}
&\int_0^\infty\f{d^2\k_1}{\pi \k_1^2}\f{d^2\k_2}{\pi\k_2^2}
\left(\f{\k_1^2}{M^2}\right)^{\g_1}
\left(\f{\k_2^2}{M^2}\right)^{\g_2}
\f{1}{[M^2+\xi(\k_1+\k_2)^2]^n}=\nn\\
&=(M^2)^{-n}\xi^{-\g_1-\g_2}
\f{\G(n+1-\g_1-\g_2)\G(1-\g_1-\g_2)\G(\g_1)\G(\g_2)}{\G(n)\G(1-\g_1)\G(1-\g_2)}.
\end{align}

The remaining integrals over the longitudinal momentum fractions and the 
Feynman parameters are evaluated in a straightforward way in terms of 
Gamma and Beta functions, 
the only trouble being the number of terms, about one hundred.
After doing the Gamma -function algebra 
the result (3.13) is found.
                           
\newpage
\appsection{Appendix B: The kernel in $\k$-space}

The inverse Mellin transform needed to compute the regular kernel (3.15) in 
transverse momentum space involves integrals of the type
\begin{equation}
\int_{\f{1}{2}-i\infty}^{\f{1}{2}+i\infty}\f{d\g}{2\pi i}x^{-\g}
\left(\f{\pi}{\sin\pi\g}\right)^2R(\g),
\end{equation}
where $x=\k^2/\k^{\prime2}$ and $R(\g)$ is a holomorphic function of $\g$, 
symmetric under the transformation $\g\ra 1-\g$, which is provided by the 
eigenvalue (3.16).

Integrals of this type can be evaluated by closing the contour on the 
l. h. poles for $x<1$ and on the r. h. ones for $x>1$, 
so that they reduce to 
infinite series. Due to the simple structure of $R(\g)$, all 
the integrals can be expressed in terms of the following basic series
\begin{align}
&\sum_{n=0}^\infty x^n=\f{1}{1-x},\\
&\sum_{n=0}^\infty \f{(-1)^nx^n}{n+\f{1}{2}}=
\f{2}{\sqrt{x}}\arctan(\sqrt{x}),\\
&\sum_{n=0}^\infty \f{(-1)^nx^n}{(n+\f{1}{2})^2}=
\f{2}{i\sqrt{x}}\left(Li_2(i\sqrt{x})-Li_2(-i\sqrt{x})\right),
\end{align}
where $Li_2$ denotes the Euler Dilogarithm.

By introducing the expression of the regular eigenvalue (3.16) in (B.1)
and performing the $\g$-integrals as described above, the expression (3.17)
for the kernel in $\k$ space follows.

\newpage
\appsection{Appendix C: Dimensional Regularization}

The peculiarity of working in $4+2\ep$ dimensions is that the kernel is 
dependent on $\ep$ as a regularization parameter, while its eigenvalues 
are finite (for $\ep\ra 0$). Therefore,
in order to compute the $q\bar{q}$ contribution to the kernel eigenvalue in 
dimensional regularization we have to compute the action of the 
($\ep$-dependent) real 
and virtual contributions on a set of test functions. By adding  
the various terms, 
the cancellation of collinear $s$-channel singularities comes out, 
and the finite $q\bar{q}$ contribution to the NL eigenvalue is obtained.
Since we want to show that the NL kernel has the form (2.1), we  
find it convenient to apply the kernel to the powerlike 
eigenfunctions of the leading kernel, i. e. 
\begin{equation}
f_\g(\k^2)=(\k^2)^{\g-1}.
\end{equation}

{\bf C.1 Real $q\bar{q}$ emission}

Let us first notice that the collinear singularities are all contained in 
the term $D_3$ in Eq. (3.10c), so that $D_1$ and $D_2$ can be computed 
directly in 4 space dimensions by taking the residue at the simple pole for 
$\g_1+\g_2=1$, along the lines of Eq. (2.25). The $D_3$ term can be evaluated 
as follows.

After a simple algebra we find
\begin{equation}
D_3=\f{2}{\nu\sh}\left[1-2(1-x_1)^2+2x_1(1-x_1)\right]+ 
\end{equation}
\begin{equation}
+\f{1}{\sh}\quad\text{or}\quad\f{1}{\sh^2}\quad\text{terms,}
\end{equation}
where in the last line we refer to  
terms of the form 
\begin{equation}
\f{P(x_1)(\k^2)^\alpha(\k^{\prime 2})^\beta}{\sh^n}
\end{equation}
(where $P(x_1)$ is a polynomial in $x_1$), 
which just vanish in dimensional regularization.

By multiplying Eq. (C.2) by the eigenfunction of the real kernel
in Eq. (C.1) and introducing the $4+2\ep$ dimensional phase-space integrals
over the internal momentum transfer $\dt$ and the transverse momentum of the
emitted $q\bar{q}$ pair, we get
\begin{align}
&D_3\otimes f_\g(\k)=\nn\\
&=(4\pi\mu^2)^{-2\ep}\as N_f\int d^{2(1+\ep)}\q d^{2(1+\ep)}\dt
\int_0^1dx_1(1-x_1)dy[(\k+\q)^2]^{\g-1}\times\nn\\
&\times\f{1+2x_1(1-x_1)-2(1-x_1)^2}{[\dt^2+x_1(1-x_1)y\q^2]^2},
\end{align}
where a Feynman parametrization of the denominator $\sim 1/(\nu\sh)$ has also 
been introduced. The integrals in (C.4) can be evaluated by first integrating 
over the two transverse momenta $\dt$ and $\q$, 
and then on the longitudinal momentum transfer $x_1$ 
and the Feynman parameter $y$.
The final result is easily expressed in terms of $\G$-functions, as follows
\begin{align}
&\bks D_3\otimes f_\g(\k)=\nn\\
&\bks=\!\left(\f{\k^2}{4\pi\mu^2}\right)^{2\ep}\!\!
\f{N_f\as}{16\pi}\f{1}{\ep^2}
\f{\G(1\!-\!\g\!-\!2\ep)\G(\g\!+\!\ep)\G(1\!+\!2\ep)\G^2(1\!+\!\ep)}
{\G(1-\g)\G(\g+3\ep)}
\left[\f{1}{\G(2+2\ep)}-\f{2(1+\ep)}{\G(4+2\ep)}\right],
\end{align}
and has the following expansion around $\ep=0$
\begin{align}
&(\k^2)^{\g-1}\left(\f{\k^2}{4\pi\mu^2}\right)^{2\ep}\f{N_f\as}{12\pi}
\left[\f{1}{\ep^2}+\f{1}{\ep}\left(2\chi_0(\g)-\f{5}{3}\right)+
2\psi^\prime(1-\g)-4\psi^\prime(\g)+\right.\nn\\
&\left.+2\chi_0^2(\g)-\f{10}{3}\chi_0(\g)
+\f{28}{9}\right].
\end{align}

Note that in $4+2\ep$ dimensions, due to the lack of scale-invariance of the 
kernel, the powers of $\k^2$ are no longer eigenfunctions of the kernel, 
and are in fact multiplied by powers of $(\k^2/\mu^2)^\ep$.
Anyway, after the combination with virtual terms, the $\ep\ra 0$ limit
will be taken, and it will be possible to isolate the non scale-invariant
terms (related to the running of $\as$), from the scale-invariant 
NL corrections. In performing this procedure we are obliged to expand all
partial results up to relative order $\ep^2$, because of the double poles 
occurring in Eq. (C.7).

{\bf C.2 Vertex corrections}

The vertex corrections to the 1-gluon production amplitude  have been 
computed in Ref. \cite{13} ($N_f$-dependent part) and \cite{11} 
($N_f$-independent part). In the following only the $N_f$-dependent 
part will be considered.

By squaring the 1-gluon production amplitude, together with its 
one-loop corrections \cite{13} 
and summing over the emitted gluon's polarization,
the contribution to the kernel in $4+2\ep$ dimensions is readily obtained, 
as follows\footnotemark
\footnotetext{Following Ref. \cite{14}, we define 
the renormalized charge in the
$\overline{MS}$ renormalization scheme as follows,
\begin{equation}
g=g_\mu\mu^{-\ep}\left[1+\left(\f{11}{3}C_A-\f{2}{3}N_f\right)
\f{g_\mu^2\G(1-\ep)}{(4\pi)^{2+2\ep}2\ep}\right].\nn
\end{equation}}
\begin{align}
&\f{N_f\as}{6\pi}\left(\f{\k^2}{4\pi\mu^2}\right)^\ep
(4\pi)^{-\ep}\G^2(1-\ep)\left[\f{1}{\ep}\f{1}{\q^2}-\f{1}{\k^2-\k^{\prime2}}
\log\f{\k^2}{\k^{\prime2}}\right]+\nn\\
&\text{terms of order $\ep$ and finite for $\q=0$}.
\end{align}

The action of this part of the (regularized) kernel on the leading order 
eigenfunctions $f_\g(\k)$ is obtained by performing the integration over $\q$
in $4+2\ep$ dimensions. However, only the first term in (C.8) needs  
be regularized, while the second one is finite in both the $\ep=0$ and 
$\q=0$ limit.

The second term is easily computed in four dimensions, by using the integral 
representation
\begin{equation}
\f{1}{\k^2-\k^{\prime2}}\log\f{\k^2}{\k^{\prime2}}=
\int_0^1\f{dy}{y\k^2+(1-y)\k^{\prime2}}
\end{equation}
and then performing the $\k^{\prime}$ and $y$ integrations according to 
the standard techniques of Appendix A, to yield the result
\begin{equation}
\f{N_f\as}{6\pi}\left(\f{\pi}{\sin\pi\g}\right)^2.
\end{equation}
The first term in Eq. (C.6) is computed by means of a Feynman parametrization 
to combine the two powers $1/\q^2$ and $[(\k+\q)^2]^{\g-1}$, and then 
performing the $2+2\ep$-dimensional integration over $\q$ 
with the standard techniques of dimensional regularization.
The final integral over the Feynman parameter is straightforward
so that the contribution of the kernel (C.6) can be expressed as
\begin{align}
&\bks-\f{1}{\ep^2}\f{\G(1-\g-\ep)\G(1+\ep)\G(\g+\ep)}{\G(1-\g)\G(\g+2\ep)
\G(1-\ep)}\f{\as N_f}{6\pi}\left(\f{\k^2}{4\pi\mu^2}\right)^\ep
\G^2(1-\ep)(4\pi)^{-\ep}+\nn\\
& -\left(\f{\as N_f}{6\pi}\right)\left(\f{\pi}{\sin\pi\g}\right)^2.
\end{align}
In the $\ep=0$ limit, (C.11) has the expansion
\begin{equation}
\bks -\f{\as N_f}{6\pi}\left(\f{\k^2}{4\pi\mu^2}\right)^\ep
\G^2(1-\ep)\left[\f{1}{\ep^2}+\f{\chi_0(\g)}{\ep}+
\f{1}{2}\left(\chi_0^2(\g)+\psi^\prime(\g)-3\psi^\prime(\g)\right)-
\left(\f{\pi}{\sin\pi\g}\right)^2\right].
\end{equation}

{\bf C.3 Trajectory correction and finiteness of the kernel:}

The purely virtual two loop contribution was computed 
in Ref. \cite{14}. The $N_f$-dependent part has the following series expansion 
around $\ep=0$
\begin{equation}
\delta^{2+2\ep}(\q)
\left(\f{\k^2}{4\pi\mu^2}\right)^\ep\f{N_f\as}{12\pi}
\left[\f{2(4\pi)^{-\ep}}{\ep^2}+
\left(\f{\k^2}{4\pi\mu^2}\right)^\ep
\left(-\f{1}{\ep^2}+\f{5}{3}\f{1}{\ep}-\f{28}{9}\right)\right].
\end{equation}

By integrating the above result over the phase-space $d^{2+2\ep}\q$ and by
combining it with the real emission contribution 
(C.7) and the vertex correction (C.12),
the action of the total NL $q\bar{q}$ kernel on the leading order 
eigenfunction $f_\g(\k)$ is finally found, as follows
\begin{align}
&K^{q\bar{q}}\otimes f_\g(\k)=\nn\\
&=\f{N_f\as}{6\pi}\left[\log\f{\k^2}{\mu^2}\chi_0(\g)+
\f{1}{2}(\chi_0^2(\g)+\chi_0^\prime(\g))-\f{5}{3}\chi_0(\g)+\right.\nn\\
&\left.-\f{1}{C_A^2}\left(\f{\pi}{\sin\pi\g}\right)^2
3\cos(\pi\g)\f{1+\f{3}{2}\g(1-\g)}{(3-2\g)(1-2\g)(1+2\g)}\right]
+O(\ep).
\end{align}

Note that only the first term is not scale-invariant, and it is 
proportional to the lowest order eigenvalue, with a coefficient which 
is just the $q\bar{q}$ contribution to the R. G. beta-function.

Finally, by combining the real emission terms in Eq. (C.7) with the virtual
ones in Eqs. (C.10), (C.12) and (C.14), the results in Eqs. (4.4) and
(4.6) of the text are obtained.

\newpage

\begin{figure}{htb}
\vspace*{1cm}
\centerline{\psfig{figure=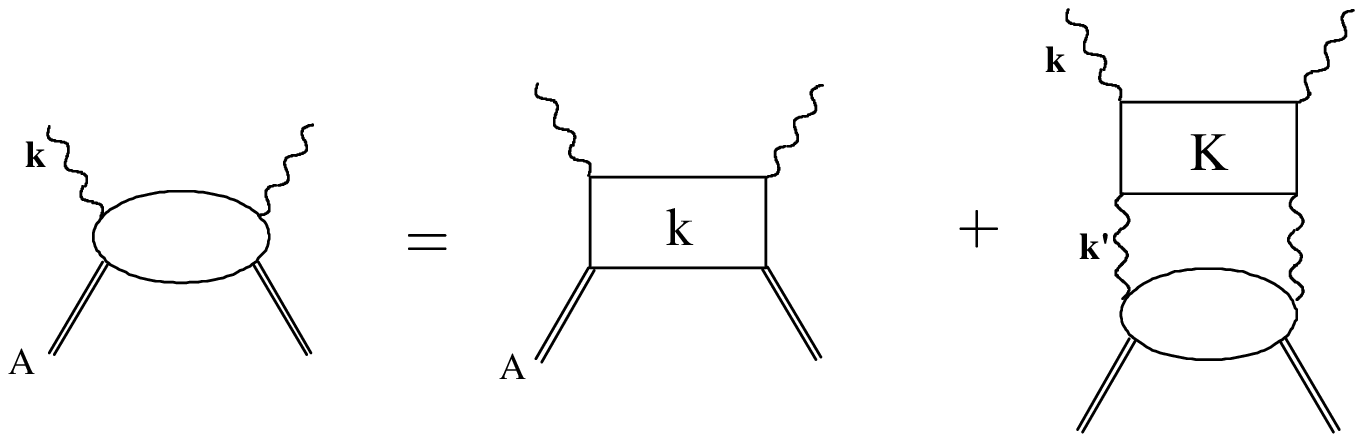}}
\vspace*{-22cm}
\caption{BFKL equation for the gluon density up to next-to-leading level. 
Wavy lines denote (Regge) gluon exchanges.}
\centerline{\psfig{figure=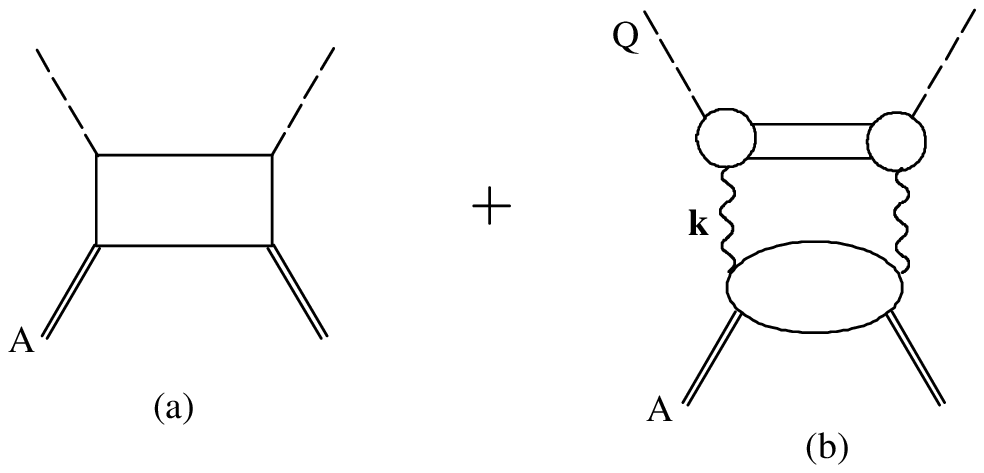}}
\caption{(a) Low energy and (b) high-energy part of the quark density in the 
DIS-$Q_0$ scheme. Dotted lines denote the electroweak probe and an $F_2$ 
projection is understood.}
\end{figure}

\newpage

\begin{figure}
\centerline{\psfig{figure=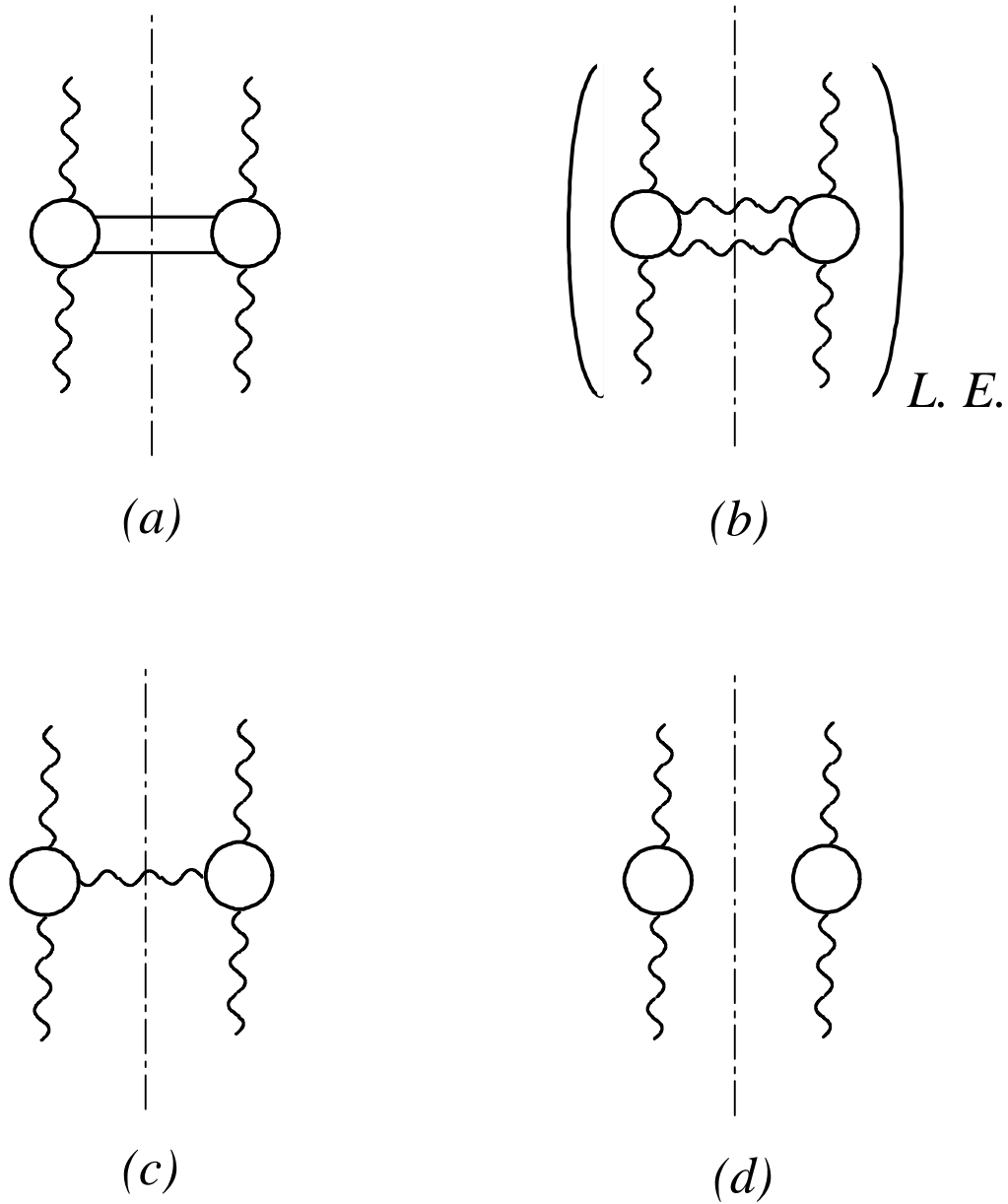}}
\caption{(a) $q\bar{q}$ and (b) low energy $gg$ contributions to the NL kernel,
together with (c) one-loop corrections to $1g$ state and (d) two-loop 
virtual corrections.}
\end{figure}

\newpage
                                          
\begin{figure}
\centerline{\psfig{figure=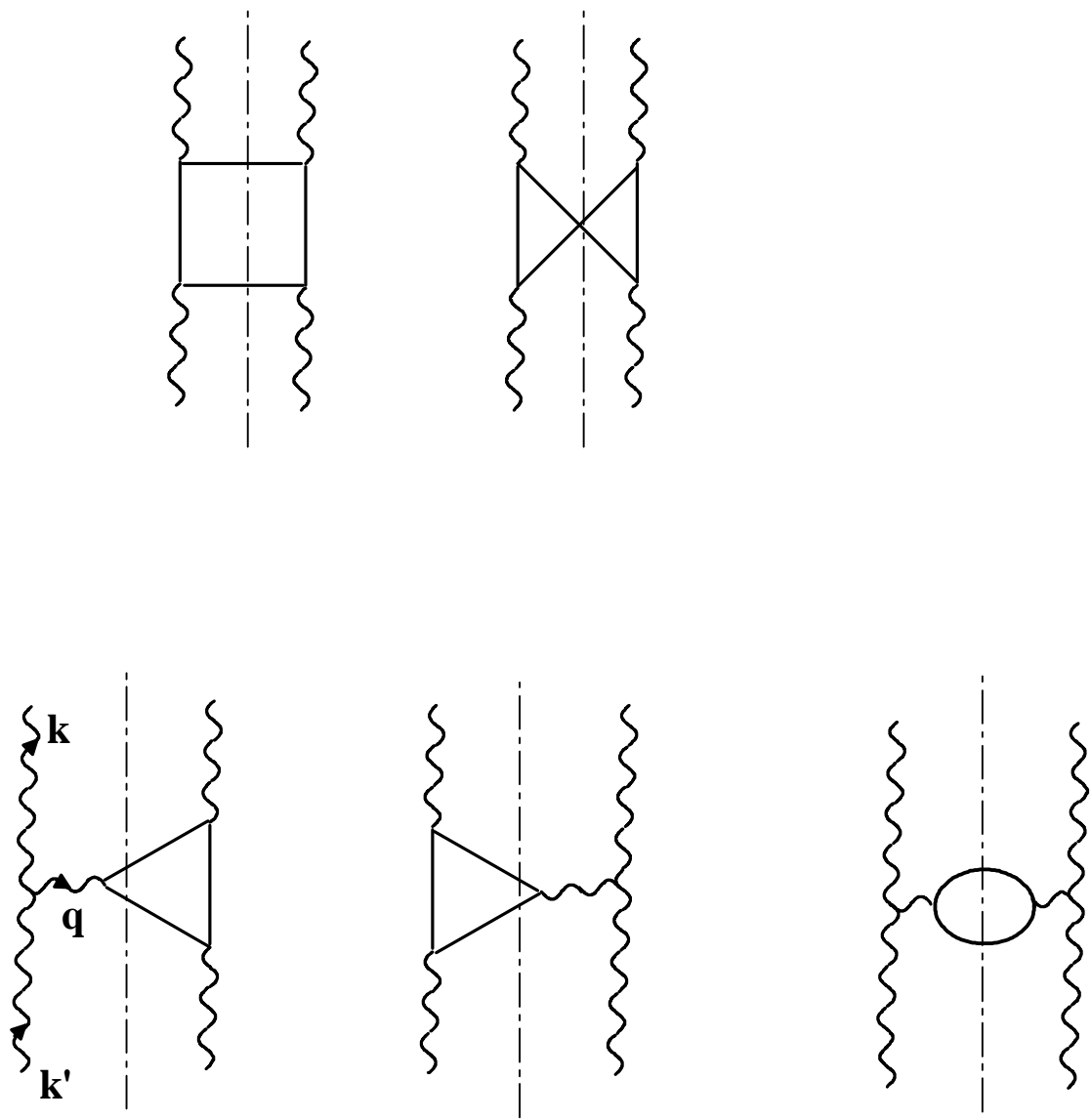}}
\caption{$q\bar{q}$ contribution to the 4 (Regge) gluon amplitude absorptive 
part.}
\end{figure}

\newpage

\begin{figure}
\centerline{\psfig{figure=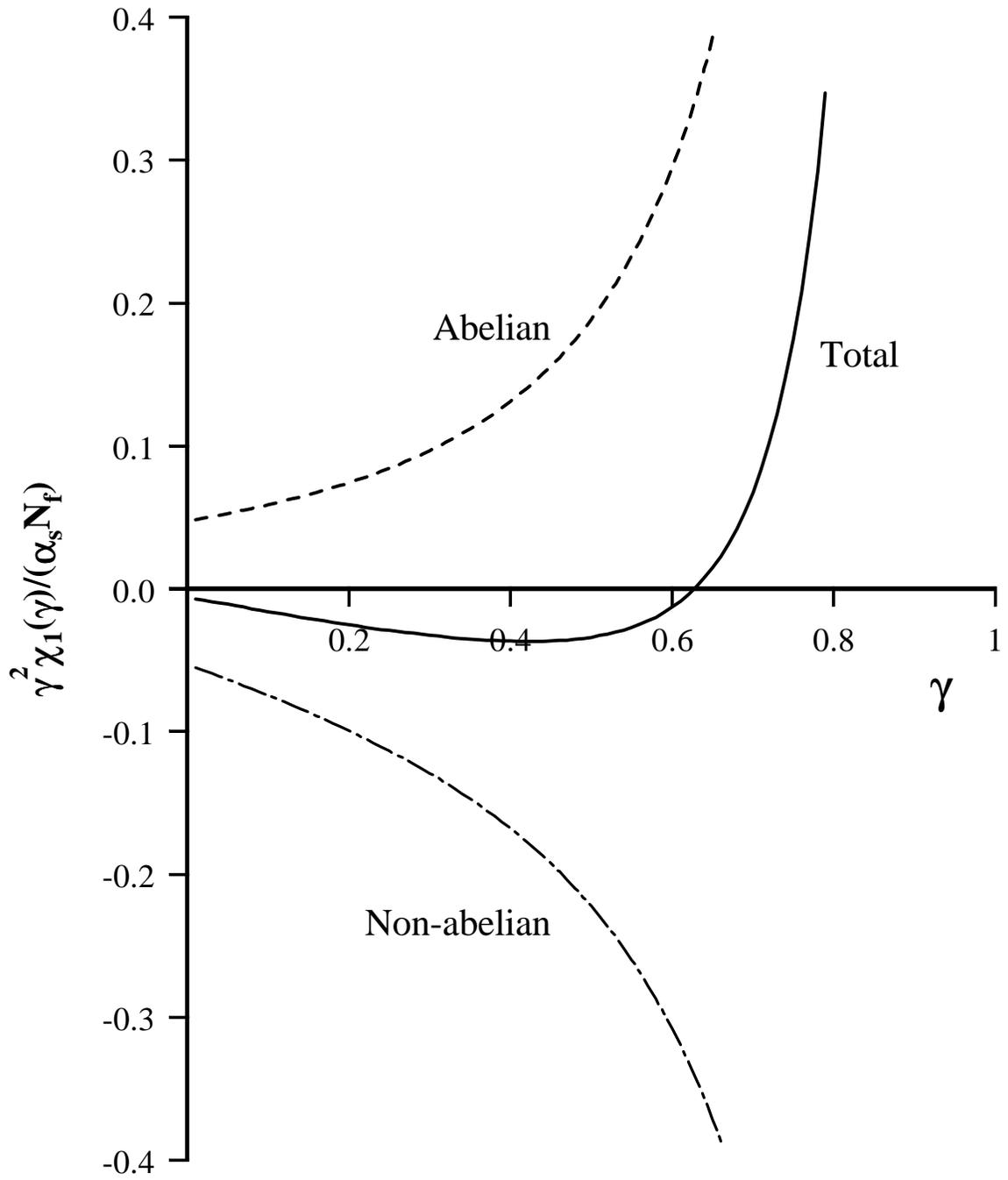}}
\caption{$q\bar{q}$ contributions to the BFKL eigenvalue as function of the
anomalous dimension variable $\g$.}
\end{figure}

\newpage 

\begin{figure}
\centerline{\psfig{figure=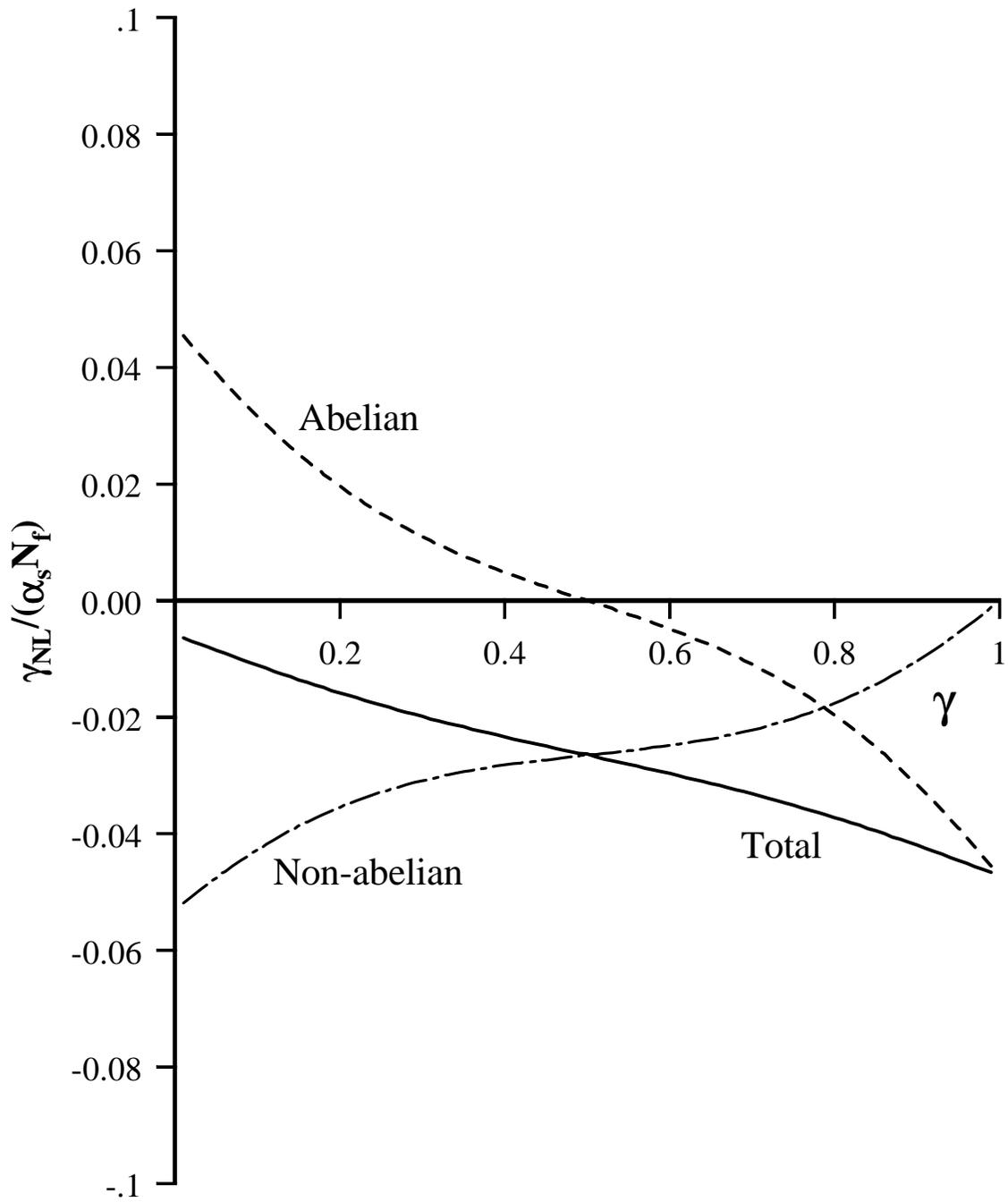}}
\caption{$q\bar{q}$ contribution to the largest eigenvalue of the 
anomalous dimension matrix $\g_+$.}
\end{figure}

\newpage

\begin{figure}{htb}
\vspace*{1cm}
\centerline{\psfig{figure=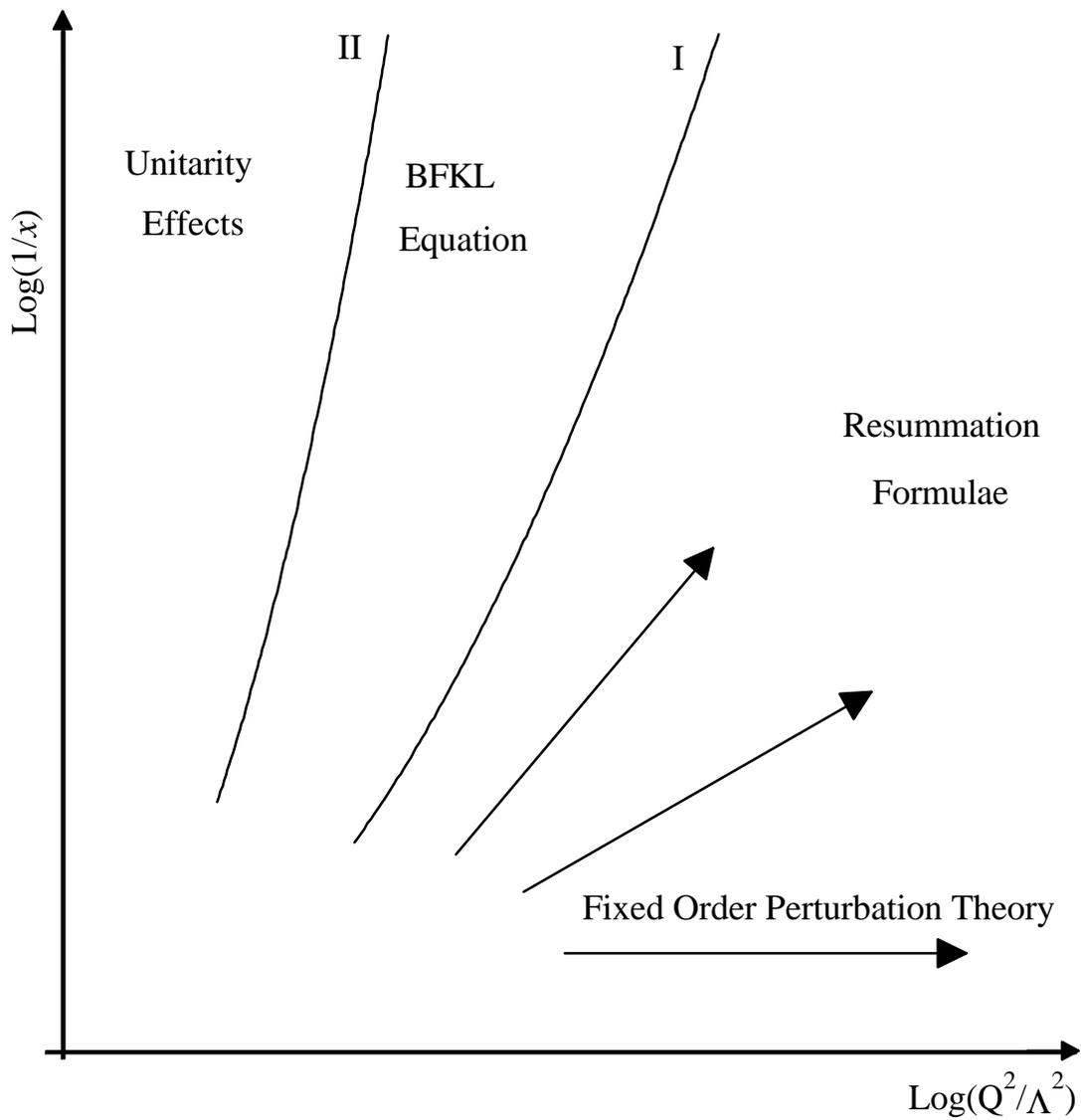}}
\vspace*{-11cm}
\caption{Qualitative plot of the anomalous dimension regimes in the 
$\log x$/$\log Q^2$ plane. Curve I is the boundary for the validity of 
resummation formulas.}
\end{figure}

\end{document}